\documentclass[aps,prx,showpacs,amsmath,amssymb,superscriptaddress,reprint,10pt]{revtex4-1}

\usepackage[colorlinks=true,breaklinks=true,allcolors=blue]{hyperref}
\usepackage{url}
\usepackage{txfonts}

\usepackage{amsmath, amssymb}

\usepackage{graphics}
\usepackage{pstricks}
\usepackage{tikz}
\usepackage{graphicx}
\usepackage{color}
\usepackage{hyperref}

\renewcommand{\i}{\mathrm{i}}

\renewcommand{\vec}[1]{\boldsymbol{#1}}
\renewcommand{\vector}[1]{\mathbf{#1}}

\newcommand{\del}{\partial}

\newcommand{\eq}[1]{(\ref{eq:#1})}
\newcommand{\Eq}[1]{Eq.\,\eqref{eq:#1}}

\newcommand{\Fig}[1]{Fig.~\ref{fig:#1}}
\newcommand{\fig}[1]{\ref{fig:#1}}

\newcommand{\Sect}[1]{Sect.~\ref{sec:#1}}

\newcommand{\App}[1]{Appendix~\ref{app:#1}}

\newcommand{\bPhase}{\theta}

\newcommand{\1}{\uparrow}

\newcommand{\eg}{\textit{e.\,g.}}
\newcommand{\ie}{\textit{i.e.}}
\newcommand{\cf}{\textit{cf.}}

\newcommand{\footnoteremember}[2]{%
\footnote{#2}%
\newcounter{#1}%
\setcounter{#1}{\value{footnote}}%
}
\newcommand{\footnoterecall}[1]{%
\footnotemark[\value{#1}]%
}

\makeatletter
\let\cat@comma@active\@empty
\makeatother

\begin{document}

\title{Strongly anomalous non-thermal fixed point in a quenched two-dimensional Bose gas}

\author{Markus Karl}
%\email{M.Karl@KIP.Uni-Heidelberg.De}
\affiliation{Kirchhoff-Institut f\"ur Physik,
             Ruprecht-Karls-Universit\"at Heidelberg,
             Im~Neuenheimer~Feld~227,
             69120~Heidelberg, Germany}
\affiliation{Institut f\"ur Theoretische Physik,
             Ruprecht-Karls-Universit\"at Heidelberg,
             Philosophenweg~16,
             69120~Heidelberg, Germany}
\affiliation{ExtreMe Matter Institute EMMI,
             GSI Helmholtzzentrum f\"ur Schwerionenforschung GmbH, 
             Planckstra\ss e~1, 
             64291~Darmstadt, Germany} 

\author{Thomas Gasenzer}
\email{T.Gasenzer@Uni-Heidelberg.De}
\affiliation{Kirchhoff-Institut f\"ur Physik,
             Ruprecht-Karls-Universit\"at Heidelberg,
             Im~Neuenheimer~Feld~227,
             69120~Heidelberg, Germany}
\affiliation{ExtreMe Matter Institute EMMI,
             GSI Helmholtzzentrum f\"ur Schwerionenforschung GmbH, 
             Planckstra\ss e~1, 
             64291~Darmstadt, Germany} 

\date{\today}

%======================================================================================
\begin{abstract}
Universal scaling behavior in the relaxation dynamics of an isolated two-dimensional Bose gas is studied by means of semi-classical stochastic simulations of the Gross-Pitaevskii model.
The system is quenched far out of equilibrium by imprinting vortex defects into an otherwise phase-coherent condensate.
A strongly anomalous non-thermal fixed point is identified, associated with a slowed decay of the defects in the case that the dissipative coupling to the thermal background noise is suppressed.
At this fixed point, a large anomalous exponent $\eta\simeq-3$ and, related to this, a large dynamical exponent $z\simeq5$ are identified.
The corresponding power-law decay is found to be consistent with three-vortex-collision induced loss.
The article discusses these aspects of non-thermal fixed points in the context of phase-ordering kinetics and coarsening dynamics, thus relating phenomenological and analytical approaches to classifying far-from-equilibrium scaling dynamics with each other.
In particular,  a close connection between the anomalous scaling exponent $\eta$, introduced in a quantum-field theoretic approach, and conservation-law induced scaling in classical phase-ordering kinetics is revealed.
Moreover, the relation to superfluid turbulence as well as to driven stationary systems is discussed.

\end{abstract}
%======================================================================================


\maketitle

%======================================================================================
\section{Introduction}
The concept of universality has been very successful in characterizing and classifying equilibrium states of matter.
It rests upon the observation that symmetries can cause the same physical laws to describe certain properties of very different systems.
For example, in the vicinity of a continuous phase transition, certain macroscopic properties of a system become independent of  the dynamical details, such as the microscopic interactions between its constituents.
In the formulation of renormalization-group theory \cite{Kadanoff1990a,Wilson1975a.RevModPhys.47.773}, universality manifests itself in scaling laws characterized by a set of universal power-law exponents.
These exponents are, to a large extent, determined by the symmetries present in the system and define which class the critical behavior belongs to.
As a result, phenomena as diverse as opalescent water under high pressure, critical protein diffusion in cell membranes \cite{Veatch2007a.PNAS104.17650}, or early-universe inflationary dynamics \cite{Hinshaw2013aApJS208.19,Ade:2013zuv} fall into the same universality class.

Taking a more general perspective, the question arises to what extent universality appears also in the dynamics away from thermal equilibrium. 
A standard classification scheme of dynamical critical phenomena applies to the linear response of systems driven, in a stochastic way, out of equilibrium \cite{Hohenberg1977a}.
Going further beyond such near-equilibrium settings, the theory of phase-ordering kinetics focuses on universal scaling laws for the coarsening dynamics following a quench across a phase transition \cite{PhysRevLett.67.2670,PhysRevE.47.R9,PhysRevE.47.228,PhysRevE.49.R27,Bray1994a,Bray2000PhRvL..84.1503B}.
Closely related dynamical critical phenomena include \mbox{(wave-)}turbulence \cite{Zakharov:Kolmogorov,Frisch1995a}, as well as superfluid or quantum turbulence \cite{Tsubota2008a, Vinen2006a}.
Universal scaling far from equilibrium has recently been analyzed for different types of quantum quenches
\cite{Braun2014a.arXiv1403.7199B,Lamacraft2007.PhysRevLett.98.160404,Rossini2009a.PhysRevLett.102.127204,DallaTorre2013.PhysRevLett.110.090404,Gambassi2011a,PhysRevB.88.201110,Smacchia2015a.PhysRevB.91.205136,PhysRevB.91.220302,Maraga2015a.PhysRevE.92.042151,Maraga2016a,Chiocchetta2016a.PhysRevB.94.134311,Chiocchetta:2016waa},
see also Refs.~\cite{Damle1996a.PhysRevA.54.5037,Mukerjee2007a.PhysRevB.76.104519,Williamson2016a.PhysRevLett.116.025301,Hofmann2014PhRvL.113i5702H,Bourges2016a.arXiv161108922B} for studies of phase-ordering kinetics in ultracold Bose gases.
While coarsening phenomena have partly been associated with the standard dynamical universality classes \cite{Bray1994a}, a rigorous renormalization-group analysis as well as a comprehensive classification scheme of far-from-equilibrium universal dynamics are lacking so far.

Here we consider possible universal scaling behavior of a time-evolving isolated two-dimensional (2D) quantum-degenerate Bose gas quenched far out of equilibrium.  
We discuss the numerically found scaling in time and in the spatial degrees of freedom in the framework of non-thermal fixed points \cite{Berges:2008wm,Berges:2008sr,Scheppach:2009wu,Berges:2010ez,Orioli:2015dxa}.
This approach builds on a scaling analysis of non-perturbative dynamic equations for field correlation functions in the spirit of a renormalization-group approach to far-from-equilibrium dynamics \cite{Gasenzer:2008zz,Berges:2008sr,Berges:2012ty,Mathey:2014xxa,Marino2016PhRvL.116g0407M,Marino2016PhRvB..94h5150M,Gasenzer:2010rq}.
Close to a non-thermal fixed point, correlation functions show a time evolution which takes the form of a rescaling in space and time  \cite{Orioli:2015dxa}.
In consequence, the relaxation is critically slowed down, while correlations evolve as a power law rather than exponentially in time.

We prepare far-from-equilibrium states by imprinting phase defects, \ie{}, quantum vortex excitations, into an otherwise strongly phase-coherent condensate.
Different kinds of initial states are realized by varying the number of defects, their arrangement, and their winding numbers.
Independently of the microscopic details of the initial state, such as the statistics of fluctuations, the system is attracted to one or more non-thermal fixed points where the information about these details gets lost.
Close to such a fixed point the correlations exhibit and evolve according to universal power laws  \cite{Nowak:2010tm,Nowak:2011sk,Nowak:2012gd,Schole:2012kt,Nowak:2013juc,Ewerz:2014tua}. 

More than one attractor can exist for the dynamical evolution of the system, as we will demonstrate being the case for the 2D Bose gas studied here. 
Consequently, different types of universal evolution with different power laws for each attractor are found. 
Nevertheless, which type of evolution is realized depends on the macroscopic properties of the initial state and on the stability properties of the attractors only.
Here we demonstrate that these macroscopic properties are given by the characteristic interaction processes within the vortex ensemble and between the defects and the fluctuating bulk \cite{Nowak:2011sk,Schole:2012kt}. 

These properties govern the scaling evolution and in particular give rise to a power-law decrease of the vortex number in time which can be described as a coarsening process  \cite{Bray1994a}. 
See also Refs.~\cite{Horng2009a.PhysRevA.80.023618,Numasato2009a,Numasato2010a.PhysRevA.81.063630,Kusumura2012a.JPhysConfSer400.012038,Reeves2012a.PhysRevA.86.053621,Reeves2013a.PhysRevLett.110.104501,White2012a.PhysRevA.86.013635,Chesler2013a.Science341.368,Neely2013a.PhysRevLett.111.235301,Kwon2014a.PhysRevA.90.063627,Stagg2015a.PhysRevA.91.013612,Moon2015a.PhysRevA.92.051601,Cidrim2016a.PhysRevA.93.033651} for theoretical and experimental studies of turbulence and vortex annihilation in a 2D superfluid.

\emph{Main result}.
Tuning the initial conditions, different non-thermal fixed points can be approached.
In this way, we identify a \emph{strongly anomalous non-thermal fixed point}, besides the standard dissipative fixed point related to coarsening of the Hohenberg--Halperin model A \cite{Bray1994a}.
At the anomalous fixed point, the coarsening process is much slower than in the standard dissipative case.
A negative anomalous scaling exponent $\eta\simeq-3$ results at the anomalous fixed point, in contrast to $\eta\simeq0$ at the dissipative fixed point.
Our results provide insight into the phenomenology behind analytical classifications of far-from-equilibrium scaling dynamics.

The anomalous fixed point is approached if the coupling of the defects to the fluctuating bulk is sufficiently suppressed.
In this case, the vortices can travel almost freely through the system.
For suitable initial conditions, vortices with equal-sign circulation tend to cluster with each other, such that the formation of closely bound pairs of vortices and anti-vortices is suppressed.
The corresponding structure factor shows a steep $k^{-6}$ momentum (Porod) law, indicating a relation to superfluid turbulence with a Kraichnan-type enstrophy cascade \cite{Schole:2012kt,Mininni2013a.PhysRevE.87.033002,Billam2014a.PhysRevLett.112.145301,Billam2015a.PhysRevA.91.023615,Kraichnan1967a}.
The resulting strongly slowed coarsening dynamics is consistent with a loss process requiring three-vortex interactions \cite{Schole:2012kt,Simula2014a.PhysRevLett.113.165302,Groszek2016a.PhysRevA.93.043614}, in contrast to the two-vortex interactions \cite{Schole:2012kt,Lucas2014a.PhysRevA.90.053617} applying in the case of dissipative coarsening \cite{Bray1994a}. 
The associated growth law  $\ell_{\text d}(t)\sim t^{1/z}$ for the characteristic length scale is found to involve a large dynamical exponent $z\simeq5$, also reminiscent of vortex glasses in type-II superconductors \cite{Nattermann2000AdPhy49607N}.

We discuss the universal dynamics near the non-thermal fixed points in the framework of the classical theory of phase-ordering kinetics.
The anomalous scaling exponent $\eta$ of the field-theoretic approach \cite{Berges:2008wm,Berges:2008sr,Scheppach:2009wu,Orioli:2015dxa} can be related to the scaling exponent $\mu$ which characterizes the coarsening dynamics of a conserved order parameter as compared to a non-conserved one \cite{Bray1994a}.
Our results are consistent with $\mu\simeq4$. 
Comparing with numerical results for a driven-dissipative Bose gas, we identify the coupling strength between the defects and the small-scale thermal fluctuations, which are seen to build up self-consistently also in the isolated system, as the relevant trigger between dissipative and anomalous scaling dynamics. 

Our article is organized as follows.
In \Sect{model} we recapitulate the Truncated-Wigner approach we use to describe the quantum field dynamics of our system, and discuss the relevant correlation functions.
\Sect{NTFP} summarizes the main aspects of non-thermal fixed points and presents our numerical results.
In \Sect{NTFPvsPHOK}, we interpret our results in the context of phase-ordering kinetics and coarsening and compare with the corresponding dynamics in a quenched-dissipative system, as well as a clustering transition and vortex-glass scaling in a driven stationary Bose gas. 
Our conclusions are drawn in \Sect{conclusions}.

%======================================================================================
\section{Non-linear Schr\"odinger model}
\label{sec:model}
The 2D Bose gas considered in this work is described by the non-linear Schr\"odinger (Gross--Pitaevskii) Hamiltonian (in the following we set $\hbar=1$),
\begin{equation}
  \label{eq:Hamiltonian}
  \text{H} = \int \! \text{d}^2x \, \left[-\frac{1}{2m}\Phi^{\dagger} \nabla^2 \Phi - \mu\Phi^{\dagger}\Phi + \frac{g}{2}\Phi^{\dagger}\Phi\Phi^{\dagger}\Phi\right]\,,
\end{equation}
where $g$ sets the interaction strength and $\mu=g\rho$ is a chemical potential, with mean density $\rho$. 
All terms are local in space and time, and we have suppressed the corresponding arguments of the bosonic field operators $\Phi$.
We focus on an isolated, homogeneous
two-dimensional system of size $L\times L$ with periodic boundary
conditions. 
We will also consider the case with coupling to a thermal bath.
Details of the description concerning the driving and dissipation relevant to that case are given in \App{DDGPE}.

In the physical situations we are studying in this work, the relevant mode occupation numbers are 
large. 
Hence, the time evolution of the quantum many-body system, described by the Hamiltonian~(\ref{eq:Hamiltonian}), can be efficiently  simulated with semi-classical numerical techniques. 
We make use of the Truncated-Wigner approximation~\cite{Blakie2008a, Polkovnikov2010a} in which
the quantum operator $\Phi(\vec{x},t)$ is replaced by a classical stochastic field variable $\phi(\vec{x},t)$ in \Eq{Hamiltonian} and the time evolution of each realization is determined by the classical Gross-Pitaevskii equation (GPE) of motion, 
\begin{equation}
  \label{eq:1}
  \i \partial_t \phi - \left[-\frac{\nabla^2 \phi}{2m} - \mu + g\phi^{\ast}\phi\right]\phi\equiv E[\phi^{*},\phi]=0\,.
\end{equation}
A state is then represented by a statistical ensemble of
fields $\phi(\vec{x},t)$ and quantum expectation values of an operator $\mathrm{O}$ are computed, in a Monte-Carlo-type fashion, via
ensemble averages with respect to a positive definite phase-space (Wigner) distribution function \cite{Berges:2007ym},
\begin{align}
  \label{eq:EnsembleAverage}
  \langle \mathrm{O}[\Phi^{\ast}(\vec{x},t),\,&\Phi(\vec{x},t)] \rangle
  = \int \mathcal{D}\phi^{*}\mathcal{D}\phi\, \mathcal{W}[\phi^{*}(\vec{x},t_\mathrm{i}),\phi(\vec{x},t_\mathrm{i})]
  \nonumber\\
  &\times\ \delta[E[\phi^{*},\phi]]\,
        O_{c}[\phi^{\ast}(\vec{x},t),\phi(\vec{x},t)]\,.
\end{align}
Here $\mathcal{W}$ is the Wigner function at the initial time $t_{\mathrm{i}}$ and $O_{c}$ is the classical (Weyl) symbol
of $\mathrm{O}$.
The delta distribution together with the functional measure 
encodes the time evolution of $\mathcal{W}$ to any time $t\not=t_{\text i}$.

As we are studying homogeneous systems with periodic boundary conditions, one of our natural observables will be the 
(angle averaged) single-particle momentum spectrum
\begin{equation}
\label{eq:occspec}
n(k,t) = \int\!\text{d}{\Omega_{k}}\,\langle\Phi^{\dagger}(\vec{k},t)\Phi(\vec{k},t)\rangle\,,
\end{equation}
where the $\Phi(\vec{k},t)$ are the bosonic field operators in momentum space, evaluated at time $t$.
Within the semi-classical approach, defining the initial state is
tantamount to defining the statistical properties of the initial
distribution of fields, $\mathcal{W}[\phi^{\ast},\phi]\vert_{t=0}$.
The initial distribution $\mathcal{W}\vert_{t=0}$ can be assumed to
describe Gaussian fluctuations of the fields, separate for every
momentum mode $\phi_k$, $\phi^{\ast}_k$. Thus, it is sufficient to
define the mean $\langle \phi^{(\ast)}_k \rangle \vert_{t_{\mathrm{i}}}$ and the
variances $\langle \phi^{(\ast)}_k \phi^{(\ast)}_k\rangle \vert_{t_{\mathrm{i}}}$
for every mode. 
Here, $\langle\cdot\rangle\vert_{t_{\mathrm{i}}}$ denotes an ensemble average with respect to the Wigner function $\mathcal{W}$, \Eq{EnsembleAverage}, at $t=t_{\mathrm{i}}$.
A mode
$k$ containing only vacuum fluctuations is described by $\langle
\phi^{(\ast)}_k \rangle \vert_{t_{\mathrm{i}}} = \langle \phi^{\ast}_k
\phi^{\ast}_k\rangle \vert_{t_{\mathrm{i}}} = \langle \phi_k \phi_k\rangle
\vert_{t_{\mathrm{i}}} = 0$, together with
\begin{equation}
  \label{eq:3}
n(k,t_{\mathrm{i}}) = \langle \phi^{\ast}_k \phi_k\rangle
\vert_{t_{\mathrm{i}}} =\frac{1}{2}\,. 
\end{equation}
The non-vanishing variance~\eq{3} accounts for the vacuum fluctuation of half a particle on average for a non-interacting mode at zero temperature, and is subtracted again before interpreting $n$ as a mode occupation.
Given the above initialization of the state the Truncated-Wigner approximation is expected to quantitatively well describe the early and intermediate-time evolution relevant in this work.

%======================================================================================
\section{Universal time evolution}
\label{sec:NTFP}
\subsection{Non-thermal fixed points in a superfluid}
The formation of turbulent ensembles of vortices in an isolated, dilute, superfluid Bose gas \cite{Nowak:2010tm,Nowak:2011sk,Nowak:2012gd} can be associated with the system approaching a non-thermal fixed point~\cite{Berges:2008wm,Berges:2008sr,Scheppach:2009wu} where the time
evolution is critically slowed down before the vortices have mutually annihilated and the system eventually
thermalizes~\cite{Schole:2012kt, Orioli:2015dxa}.  Turbulent behavior and critical slowing down are 
signaled by characteristic scaling properties of correlation functions such as the angle-averaged
single-particle momentum spectrum defined in \Eq{occspec} 
\footnote{We use the term `critical slowing down' interchangeably with dynamical scaling behavior near a non-thermal fixed point.}.

An initial overpopulation \cite{Berges:2012usb,Nowak:2012gd} of momenta at scales on the order of the inverse healing length
$k_{\xi}=\sqrt{2g\rho m}$, induced, \eg{}, by an instability, can subsequently drive the system to a
non-thermal fixed point \cite{Berges:2015kfa}.  
This happens because the particles in the overpopulated region are transported in momentum space to modes with lower energy. 
Close to a non-thermal fixed point, this inverse transport is generically self-similar in space and time, obeying the
scaling relation~\cite{Orioli:2015dxa}
\begin{equation}
\label{eq:NTFPScaling}
n(k,t) = (t/t_{0})^{\alpha} n([t/t_{0}]^{\beta}k,t_{0}).
\end{equation}
At the same time, energy conservation forces a few particles to scatter into higher momentum modes, eventually forming an incoherent thermalized fraction of the gas.

The scaling exponents $\alpha$ and $\beta$ in \Eq{NTFPScaling} have been calculated for the non-linear Schr\"odinger model \cite{Orioli:2015dxa}. 
The predictions are derived by means of a scaling analysis of Dyson-type field dynamic equations obtained with $2$-particle-irreducible effective-action techniques \cite{Berges:2015kfa}.

Assuming a quasiparticle description to apply, these equations lead to effective quantum Boltzmann equations, incorporating a non-perturbative momentum-dependent many-body scattering matrix. 
A crucial role is taken by the scaling of this scattering matrix.
It is obtained within a non-perturbative scheme, summing diagrams to all orders in the bare coupling $g$ \cite{Berges:2008wm,Berges:2008sr,Scheppach:2009wu,Berges:2010ez}, equivalent to a next-to-leading order large-$\mathcal{N}$ approximation for the case of an $O(\mathcal{N})$-symmetric model.

The effective quantum Boltzmann equations take into account the scattering of quasiparticle modes the properties of which are encoded by a spectral function $A_{ij}(\omega,\vec k)$.
During the scaling evolution, this function is assumed to be stationary.
It can thus be written as the expectation value of the commutator matrix of the bosonic 
field operators 
\begin{equation}
\label{eq:Aij}
A_{ij}(\omega,\vec{k}) =
\left\langle\left[ \Phi_{i}(\omega,\vec{k}),\, \Phi^{\dagger}_{j}(\omega,\vec{k})\right]\right\rangle\,,
\end{equation}
$i,j\in\{1,2\}$, where $\Phi_{1}(\omega,\vec{k})\equiv\Phi(\omega,\vec{k})$, and $\Phi_{2}(\omega,\vec{k})\equiv\Phi^{\dagger}(-\omega,-\vec{k})$, defining the frequency dependent response of eigenmode $\vec k$.
At the non-thermal fixed point, the spectral function is assumed to scale as
\begin{equation}
\label{eq:AijScaling}
A_{ij}(\omega,\vec{k}) = s^{2-\eta}A_{ij}(s^{-z}\omega,s\vec{k})\,.
\end{equation}
$z$ denotes the dynamical critical exponent.
For $d$-dimensional systems close to their upper critical dimension, $d\lesssim d_{\text{up}}$, the anomalous exponent $\eta$ is expected to be small.
Note, however, that $\eta$ at a non-thermal fixed point does not need to be equivalent to the static anomalous dimension at an \emph{equilibrium} phase transition.
Below, we present an example, in $d=2$, where $\eta$ is distinctly different from zero, related to an anomalously small temporal scaling exponent $\beta$ in \Eq{NTFPScaling}. 

For the GP model \eq{Hamiltonian}, the exponents $\alpha$ and $\beta$ read  \cite{Orioli:2015dxa} 
\begin{align}
  \label{eq:2}
  \alpha &= d\,\beta\,,\nonumber\\ 
  \beta&=1/(2-\eta)\,,  
\end{align}
under the condition of particle number conservation, 
\begin{equation}
\label{eq:ParticleConserv}
 \rho\sim\int_{0}^{\infty}\mathrm{d}k\,k^{d-1}n({k},t) \equiv \text{const.}\,,
\end{equation}
within the region of self-similar transport 
\footnote{$\alpha=(d+z)\,\beta$ would apply, \eg{}, in the case of a self-similar transport conserving energy, $\epsilon=\int_{0}^{\infty}\mathrm{d}k\,k^{d-1}\omega(k)\,n({k},t) \equiv \text{const.}$, for a dispersion law obeying the scaling $\omega(sk)=s^{z}\omega(k)$.}.
Setting the anomalous dimension $\eta$ to zero, one obtains the set of exponents 
$\alpha=d/2$, and $\beta=1/2$ which have been numerically confirmed for the (three-dimensional) Bose 
gas in Ref.~\cite{Orioli:2015dxa}.

If $\beta>0$ the self-similar build-up \eq{NTFPScaling} of momentum occupations in the infrared reflects a coarsening process.
This coarsening can be interpreted in terms of the dynamics and annihilation of vortex defects created through the initial quench \cite{Nowak:2011sk, Schole:2012kt}.
The vortices are moving through the superfluid gas, interacting with each other and with background sound waves.
This allows for a dilution of the defects through mutual annihilation of vortices with opposite circulation~\cite{Nowak:2011sk, Schole:2012kt}. 
The coarsening evolution implies an algebraic growth law, 
\begin{equation}
\label{eq:GrowthLaw}
\ell_{\text d}(t) \sim t^{\,\beta}\,,
\end{equation}
for the characteristic mean spacing $\ell_{\text d}$ between the vortices.
Eventually, when the last defects have disappeared, long-range order is established and phase coherence
of the Bose-condensed gas is maximized.

In the intermediate-wavelength region of the inverse transport, $k\gg1/\ell_{\text d}$, the
momentum distribution is found to exhibit characteristic scaling
$n(k) \sim k^{-\zeta}$ in momentum space \cite{Berges:2008wm,Berges:2008sr,Scheppach:2009wu,Berges:2010ez},
\begin{equation}
\label{eq:58}
n(k,t) \sim k^{-d-2+\eta}\,,
\end{equation}
\eg{}, $n_{k}\sim k^{-4}$ in $d=2$ dimensions and for vanishing $\eta$.
For $\eta\leq2$, this power-law behavior is cut off in the infrared at a certain scale $k_{\lambda}$ to
ensure non-divergence of the overall particle density $\rho$, \cf{}~\Eq{ParticleConserv}.  
For the case of vortices in a dilute Bose gas, $d\in\{2,3\}$ the power law \eq{58}, with $\eta=0$, reflects the geometry of the superfluid circulation around the defects, implying a characteristic power-law fall-off of the velocity, $v\sim|\nabla\mathrm{arg}\phi|\sim1/r$, at distance $r$ perpendicular to the vortex (line).
The scaling \eq{58} is also well known (for $\eta=0$) as the Porod tail  \cite{Bray1994a,poroda} of the momentum distribution of an ensemble of randomly distributed vortices, at $k>k_{\lambda}$, where the scale $1/k_{\lambda}$ is related to the mean distance between defects \cite{Bray1994a,Nowak:2011sk}. Here, we introduce the definition \eq{58} as a natural extension of Porod's law to include an anomalous dimension $\eta$ (see \Sect{NTFPPorodTails} for a more detailed discussion).  

%==================================================
\begin{figure}[!t]
\includegraphics{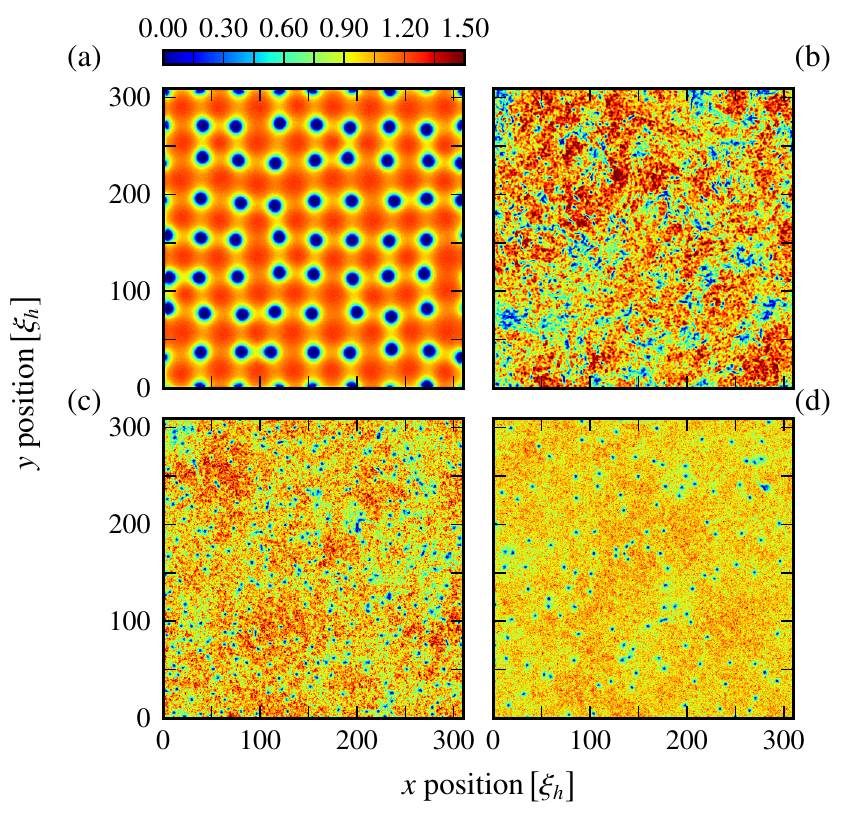}
\caption{Time evolution of a 2D superfluid in a square 
  volume with periodic boundary conditions, starting from a
  lattice of non-elementary vortices with alternating winding numbers $w=\pm6$,
  arranged in a checker-board manner
  (panel (a)).
  Color encodes the superfluid density normalized to the mean density,
  $\lvert\phi\rvert^2/\rho$.
  The vortices quickly break up into clusters of elementary defects with $w=\pm1$.
  This decay is enhanced by the slight initial displacements of the defects
  with respect to the square lattice.
  Panels (b)--(d) show snapshots of the evolving density at times
  $t=\left\{300,10^3,10^4\right\}\,\xi_h^2$.  
  In the early non-universal stage of the time evolution (panel (b)), 
  the vortex configuration is strongly clustered.
  During the later stages shown exemplarily in (c) and (d), the vortices and 
  anti-vortices mutually annihilate in a way that the vortex number falls off
  in a universal manner as $N(t)\sim t^{-2/5}$ (see main text).}
   \label{fig:vorlat}
\end{figure}
%
%==================================================
%==================================================
%
\begin{figure}[!t]
  \includegraphics[]{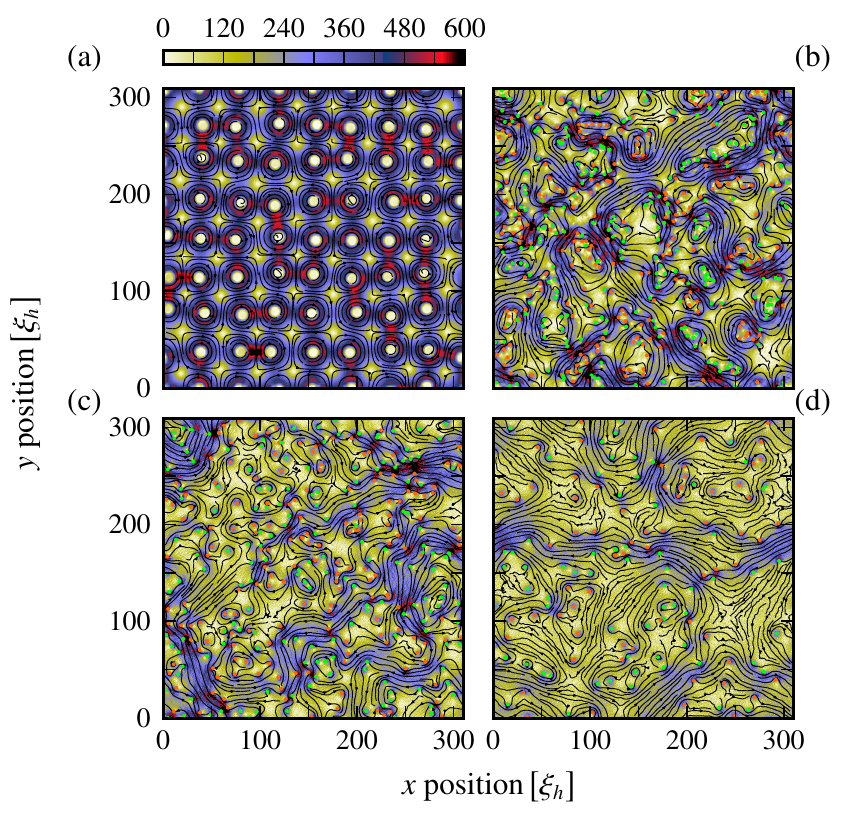}
  \caption{%
    The hydrodynamic velocity field $\vec{v}$ corresponding to the
    snapshots of the time evolving density shown in \Fig{vorlat}.
    Black flow lines indicate the orientation of
    the velocity field while the color projection depicts the modulus
    $\lvert \vec{v} \rvert$. The positions of elementary vortices (anti-vortices)
    in panels (b)--(d) are marked by orange (green) dots. 
    The strong clustering after the initial break-up as well as the large-scale
    vortex clustering leading to strong coherent flows during the universal,
    late-time stage are clearly seen.
    For comparison we provide corresponding flow diagrams for the 
    evolution near the dissipative (Gaussian) fixed point in \App{SpatialPatterns}.
    }
  \label{fig:flowlines}
\end{figure}
%==================================================

%======================================================================================
\subsection{Preparation of the initial state}
\label{sec:VortIni}
In this work, we demonstrate that different non-thermal fixed points can be reached, depending on the particular choice of the initial non-equilibrium state.
Universality implies that the dynamical evolution in the vicinity of the fixed point is insensitive to details of the initial state. 
On the other hand, it is reasonable that the initial state needs to fulfil certain conditions for the system to enter a universal regime of time evolution in the first place. 

We generate initial states by
phase-imprinting vortex defects into a fully phase-coherent Bose gas \cite{Chesler2013a.Science341.368,Ewerz:2014tua}.
This offers different parameters, such as the density of vortices, their winding number, and the distribution statistics, to vary the initial conditions. 

A fully phase-coherent gas is prepared, given by homogeneous field configurations.
In each sample prepared within the Truncated-Wigner scheme, quantum fluctuations are included in the empty modes. 
We choose, in particular, the mode occupation numbers according to \Eq{3}, for all modes $k\neq0$, while the zero mode is populated with the total number of particles $N$.
Then, the phase pattern $\bPhase(\vec{x},t_{\text i})$ of the desired vortex distribution is multiplied into the sampled homogeneous field configurations, $\phi(\vec{x},t_{\text i}) \to \phi(\vec{x},t_{\text i})\cdot \exp\{\i\,\bPhase(\vec{x},t_{\text i})\}$. 
We study two types
of initial vortex configurations, regular lattices of non-elementary defects with
winding numbers $|w| > 1$, and uniform random distributions of elementary defects
with winding number $w = \pm 1$. For both types, we choose
distributions with equal numbers of vortices and anti-vortices. 
Vortex cores in the density $\lvert \phi(\vec{x},t_{\text i})\rvert ^2$ are subsequently formed by means of a short period of evolution according to the imaginary-time GPE, \ie{}~with $t \to -\i t$ in \Eq{1}. 
In \Fig{vorlat}a, an example for such a lattice configuration with slightly displaced ($w = \pm 6$)-vortices is shown, depicting the superfluid density field $\lvert\phi(\vec{x},t_{\text i}) \rvert ^2$ after imaginary time evolution.
The random displacements, in addition to the Truncated-Wigner noise, speed up their decay into elementary vortices.
All numerical data is shown in units expressed in terms of the healing length $\xi_h = 1/\sqrt{2m\rho g}$
(with mass $m=1/2$ set in the numerical calculations), see \App{DynamicalSimulations} for more details.

%==================================================
\begin{figure}[!t]
\centering
  \includegraphics{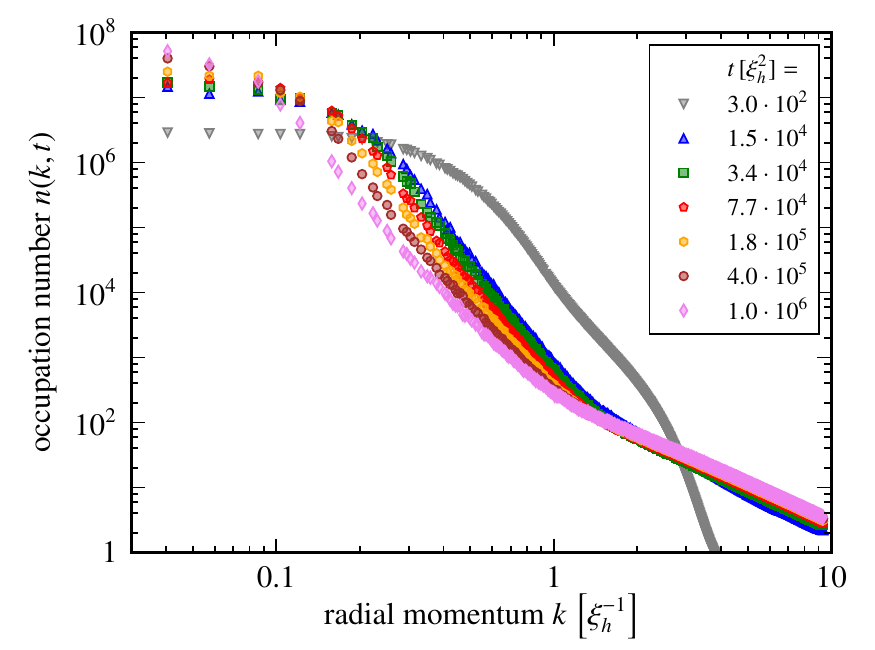}
  \caption{Single-particle momentum spectrum at seven,
    logarithmically equidistant times starting from the
    initial configuration bearing an ($8 \times 8$)-lattice of non-elementary
    vortices with winding number $w = \pm6$, as exemplary shown
    in Figs.~\fig{vorlat} and \fig{flowlines}.
    After the initial build-up of a characteristic distribution, a self-similar shift towards smaller momenta is seen during the universal
    stage of the time evolution.
    This inverse transport process conserves particle number and, as compared
    to a turbulent cascade, is non-local in momentum space. 
    The earliest time shown
    ($t=3\cdot10^{2}\,\xi_h^2$, grey points) is still within the
    non-universal stage of evolution right after the decay of the
    lattice (\cf{} \Fig{vorlat}b). 
    The dynamics during the later stage of the evolution 
    represents a rescaling in space and time and is analyzed in more detail in \Fig{vorlat_spectrum}.}
  \label{fig:vorlat_spectrum_unresc}
\end{figure}
%==================================================

%==================================================
\begin{figure}[!t]
\centering
  \includegraphics{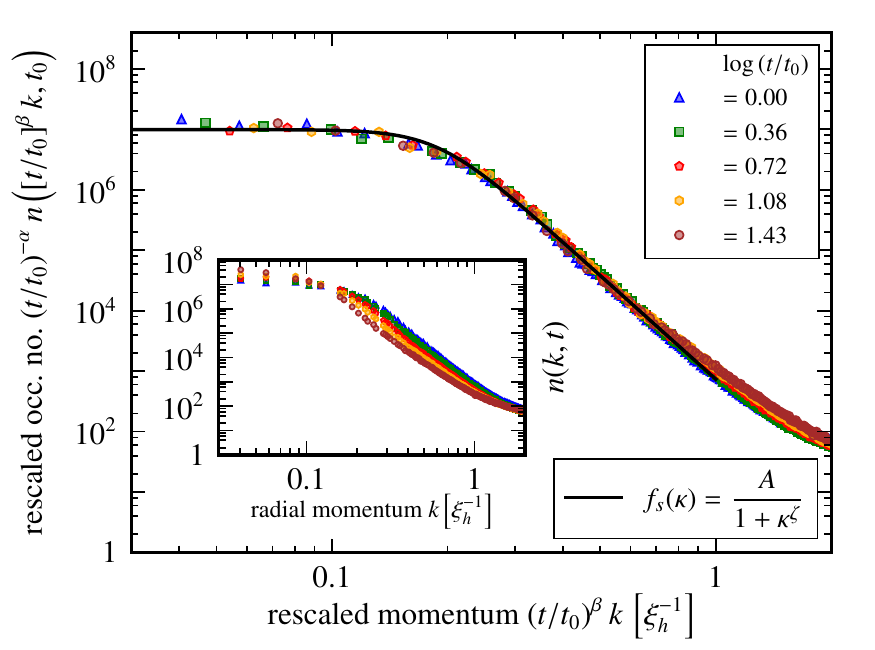}
  \caption{%
    Scaling analysis of the time evolving single-particle spectra shown in
    \Fig{vorlat_spectrum_unresc}, exhibiting the scaling at the anomalous 
    non-thermal fixed point. 
    The inset shows the non-rescaled spectra at five different, logarithmically equidistant times as given
    in the legend.
    The times are defined relative to the reference time $t_0 = 10^4\,\xi_h^2$, such that the latest time
    is $t=t_{0}\cdot10^{1.43}=2.7\cdot10^{5}\,\xi_h^2$.
    The main graph depicts the scaling collapse of
    the spectrum according to $n(t,k) =
    (t/t_0)^{-\alpha}n([t/t_0]^{\,\beta} k,t_{0})$. 
    The respective scaling exponents are determined by a fit which yields 
    $\alpha = 0.402\pm0.05$ and
    $\beta = 0.193\pm0.05$.  
    The functional form \eq{4} is fit to the scaling function 
    (black line) and shows a Porod-tail
    scaling exponent $\zeta = 5.7\pm0.3$.  
    For a better visibility we show only five
    exemplary curves, while the scaling-collapse fits
    involve the whole data set, consisting of
    spectra at $300$ logarithmically equidistant times within the
    universal stage of the evolution.
    }
  \label{fig:vorlat_spectrum}
\end{figure}
%==================================================

%======================================================================================
\subsection{Relaxational production of clustered vortex ensembles}
\label{sec:SpatialPatternEvolution}
Before analyzing the dynamics of occupation numbers ensuing the phase
imprinting, we discuss the spatial dynamics on
phenomenological grounds, at the level of single realizations of the
order parameter field $\phi(\vec{x},t)$. We focus on an initial
arrangement of the vortex defects on a rectangular
lattice with alternating signs of the winding number $w = \pm6$, see \Fig{vorlat}a. 
During the early time evolution, these defects quickly decay into
elementary vortices with $|w|=1$ \cite{PethickSmith}.
To speed up this otherwise rather slow decay process, we add initially small random shifts to the regular lattice positions
of the defects  (see \Fig{vorlat}a).

In a first stage, the like-sign elementary vortices form tightly-bound clusters, see \Fig{vorlat}b,
which play an important role in the ensuing dynamics.
Each cluster screens the vortex--anti-vortex attraction \cite{Onsager1949b}, 
such that mutual annihilation of vortices and anti-vortices is suppressed.
\Fig{flowlines} shows examples of the hydrodynamic velocity field $\vec{v}(\vec{x},t) = \nabla 
\bPhase (\vec{x},t)$ at the same times $t$ as in \Fig{vorlat}. 
The clustering of like-sign vortices is indicated by (color encoded) extended and strong fluxes.

In later stages, vortices with opposite winding numbers begin to mingle
with each other, overcome the screening and start to mutually
annihilate. As a consequence, the vortex configuration undergoes a
dilution process, \cf{} \Fig{vorlat}, panels (c) and (d), which
leads to an ordering of the phase field \cite{Nowak:2010tm,Nowak:2011sk,Nowak:2012gd,Schole:2012kt}.
We discuss the universal character of this ordering process in the
next section.

As seen in \Fig{vorlat}b--d, the superfluid density of a single
realization develops a considerable amount of small-scale fluctuations. These
fluctuations arise in the decay of the non-elementary vortices as well as in the mutual annihilations and are amplified by the
non-linear interaction term in \Eq{1}. They play an important role for the dilution
process, as they are known to mediate vortex
interactions~\cite{Damle1996a.PhysRevA.54.5037,Lucas2014a.PhysRevA.90.053617}.

%==================================================
\begin{figure}[!t]
\centering
  \includegraphics{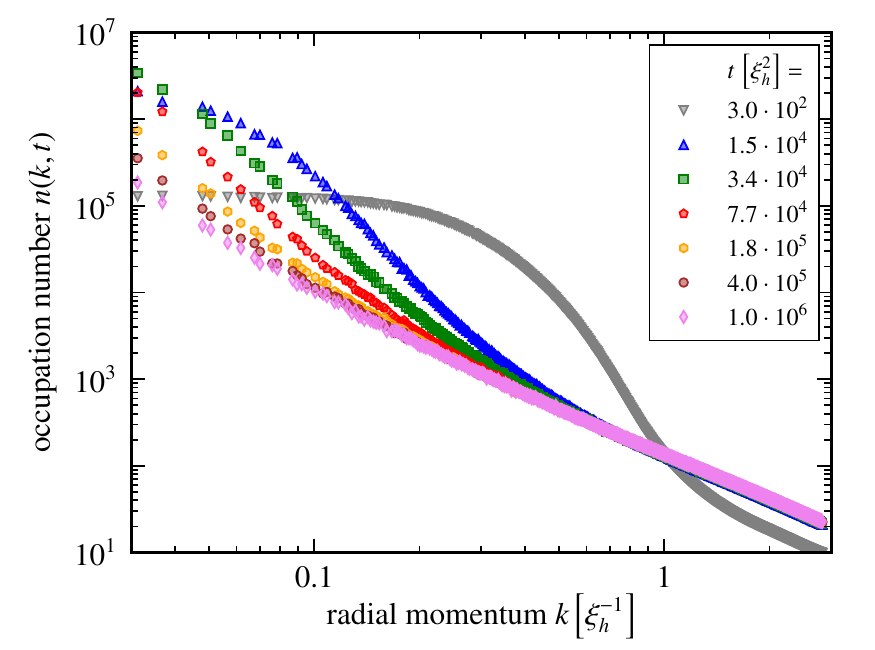}
  \caption{Single-particle momentum spectrum at seven,
    logarithmically equidistant times starting from the
    initial configuration bearing a random distribution of $2400$ elementary
    vortices with winding number $w = \pm1$, summing to a total
    angular momentum zero.
    After the initial build-up of a characteristic distribution, a self-similar 
    shift towards smaller momentum scales is seen during the universal
    stage of the time evolution.
    The earliest time shown
    ($t=3\cdot10^{2}\,\xi_h^2$, grey points) is still within the
    non-universal stage. 
    The scaling during the later universal stage
    is analyzed in \Fig{ranvort_spectrum}.
  }
  \label{fig:ranvort_spectrum_unresc}
\end{figure}
%==================================================

%=======================================================================================
\subsection{Anomalous non-thermal fixed point}
Near a non-thermal fixed point, the time evolution of the single-particle spectrum $n(\vec k,t)$, \Eq{occspec}, is expected to correspond to a scaling transformation \eq{NTFPScaling} of a universal scaling function.  
The occupation spectrum corresponding to the evolution starting from the vortex-lattice state described in the previous section is shown in \Fig{vorlat_spectrum_unresc}.
After an initial stage during which the non-elementary vortices decay, a power-law dependence of $n$ on $k=|\vec k|$ builds up in an infrared momentum region.
This steep power law levels off at a scale $k_\lambda$ in the infrared. 
This scale shifts in time towards lower momenta increasing the occupation in the infrared, while the occupation decreases at larger wave numbers. 
The evolution represents self-similar particle transport towards the infrared, building up a quasi condensate at low wave numbers, similarly as described for the 3D case in Refs.~\cite{Svistunov1991a,Kagan1992a,Kagan1994a}.

The self-similar nature of the time evolution of $n(k,t)$, depicted in
\Fig{vorlat_spectrum_unresc}, is demonstrated by the scaling collapse
according to \Eq{NTFPScaling}. \Fig{vorlat_spectrum} shows
that the time evolution of the occupation spectrum can indeed be
rescaled to a single curve. 
The fitting procedure which is described in detail in \App{fitting} yields the scaling exponents
\begin{align}
  \label{eq:82}
\nonumber  \alpha_{\text{a}} = 0.402\pm0.05\,,\\
  \beta_{\text{a}} = 0.193\pm0.05\,.
\end{align}
The index `a' is chosen to distinguish the exponents from those obtained later with different
initial conditions.
In \Fig{vorlat_spectrum}, the rescaled spectra are shown at five exemplary times between $t_0=10^4\,\xi_h^2$ and
time $t=2.7\cdot10^{5}\,\xi_h^2$, $\log(t/t_{0})=1.43$.
For the scaling analysis we took into account the occupation spectra at $300$ logarithmically equidistant times within the same interval.

The scaling evolution  shown in \Fig{vorlat_spectrum_unresc} can be interpreted to reflect
turbulent (non-local inverse particle) transport. 
We note that this inverse transport does not conserve particle number locally, \ie{}, momentum shell per momentum shell, as 
commonly required for (wave) turbulent cascades within an inertial interval of wave numbers.
Nevertheless, in the isolated system we consider, the transport conserves the total particle number.
%==================================================
\begin{figure}[!t]
\centering
\includegraphics{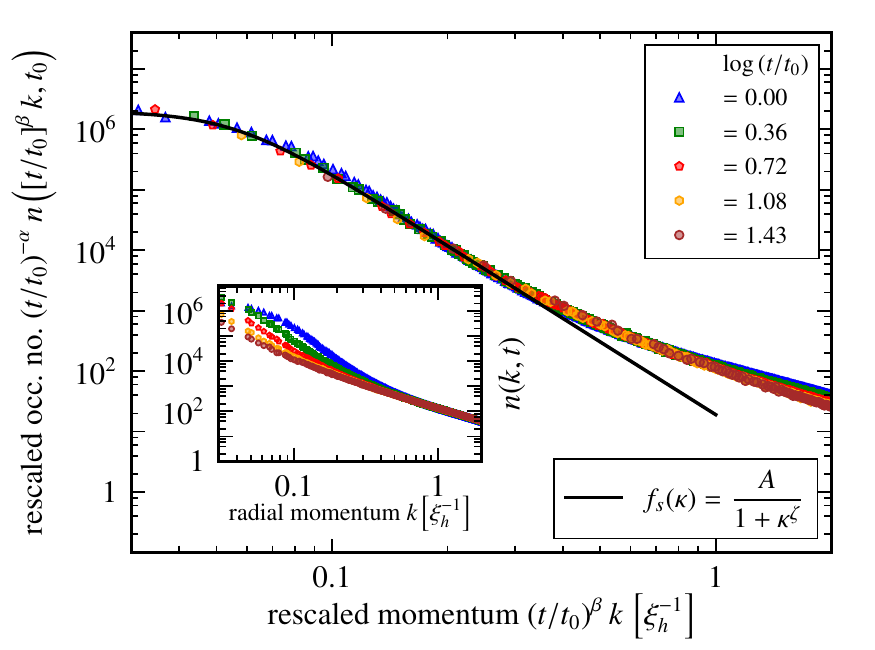}
  \caption{The graphs illustrate the scaling analysis of the evolution close to the
    Gaussian non-thermal fixed point as shown in
    \Fig{ranvort_spectrum_unresc}, in full analogy with \Fig{vorlat_spectrum}.
    The reference time is, again, $t_0 = 10^4\,\xi_h^2$. 
    The scaling exponents are determined to be $\alpha = 1.10\pm0.08$
    and $\beta = 0.56\pm0.08$.  The scaling function (black line) 
    exhibits a Porod-law exponent $\zeta = 4.0\pm0.1$.  
    }
  \label{fig:ranvort_spectrum}
\end{figure}
%==================================================

Our simulations give scaling exponents $\alpha$ and $\beta$ consistent with the relation $\alpha=d\beta$, \cf{}~\Eq{2}. 
Hence, particle number conservation also holds within the regime of momenta and times governed by the scaling evolution shown in \Fig{vorlat_spectrum}, whereas a different relation between $\alpha$ and $\beta$ would apply if the self-similar transport conserved energy or other quantities \cite{Orioli:2015dxa}.
Based on the predictions \eq{2} for the scaling exponents $\alpha$ and $\beta$
we infer a strong anomalous scaling exponent 
\begin{equation}
  \label{eq:etaanomalous}
  \eta_\mathrm{a} \simeq -3
\end{equation}
to characterize the non-thermal fixed point reached from our vortex-lattice 
initial conditions.

%=======================================================================================
\subsection{(Near-)Gaussian fixed point}
In contrast to the above anomalous scaling, characterized by the exponents \eq{82} and \eq{etaanomalous}, also a distinctly different self-similar evolution is possible when starting from different initial states.
In the following, we present results for initial states containing random spatial distributions of $2400$ elementary defects, with an equal  number of vortices and anti-vortices.
This corresponds to four times as many vortices as obtained from the decaying non-elementary vortices chosen in the
previous section.

In \Fig{ranvort_spectrum_unresc}, the time evolution of the occupation spectrum is shown in analogy
to \Fig{vorlat_spectrum_unresc}.
\Fig{ranvort_spectrum} shows the scaling collapse within the same time frame as
\Fig{vorlat_spectrum} in the anomalous case. 
As before, at late times, $t\gg10^4\,\xi_h^2$, the spectrum evolves in a universal
self-similar way. The scaling exponents are 
\begin{align}
  \label{eq:alphabetag}
\nonumber  \alpha_{\text{g}}=1.10\pm0.08\,,\\
  \beta_{\text{g}}=0.56\pm0.08\,.
\end{align}
which is again consistent with particle conservation, $\alpha=d\,\beta$, \cf{}~\Eq{2}. 
The crucial difference to what we found in the previous section is that, 
here, the numerically determined exponents are consistent with the analytical
prediction, \Eq{2}, for a vanishing anomalous exponent, indicating a (near) Gaussian
non-thermal fixed point:
\begin{equation}
  \label{eq:etagaussian}
  \eta_\mathrm{g} \simeq 0
\end{equation}
The index `g' refers to near-Gaussian, in contrast to the strongly anomalous scaling (`a') found in the
previous section.
We emphasize that our findings are compatible with a small but non-zero anomalous scaling \cite{Schachner:2016frd}. 
When we refer to the fixed point as the `Gaussian' one in the following, it is meant in relation to 
the much larger value \eq{etaanomalous} at the anomalous fixed point, not in an absolute sense.
This nomenclature includes our anticipation that a small non-zero $\eta_{\text{g}}$ renders the fixed point to be of the Wilson-Fisher type.

%======================================================================================
%
\section{Non-thermal fixed points and phase-ordering kinetics}
\label{sec:NTFPvsPHOK}
As pointed out in the previous section, if the scaling exponent $\beta$ is positive, the self-similar build-up \eq{NTFPScaling} of momentum occupations in the infrared reflects a coarsening process.
In the following, we discuss in more detail the scaling near the observed non-thermal fixed points in the context of the classical theory of phase-ordering kinetics \cite{PhysRevLett.67.2670,PhysRevE.47.R9,PhysRevE.47.228,PhysRevE.49.R27,PhysRevLett.62.2841,cardy1992a} (see Ref.~\cite{Bray1994a} for a review). 
For this, we compare the spatio-temporal scaling relation \eq{NTFPScaling} at the
non-thermal fixed point with the scaling forms obtained for the coarsening of
ensembles of defects.

Although the evolution of the Bose field, in our settings, is not described by a diffusion equation, as usually is the case in phase-ordering kinetics, we do not expect the effective coarsening dynamics to be fundamentally different in character.
As far as scaling laws are determined through the geometric and topological properties of the order-parameter field, the theory of phase-ordering kinetics is expected to apply in principle.
In fact, the dynamics at the non-thermal fixed point is found to be in the classical-wave limit of the underlying quantum dynamics.
We point that, nevertheless, the scaling exponents resulting for the coarsening dynamics of the unitarily evolving quantum system can be outside the commonly considered universality classes.
Not at last, non-thermal fixed points are expected to be the underlying principle for a larger class of universal dynamics phenomena, beyond the closer realm of phase-ordering kinetics.

%=======================================================================================
\subsection{Porod tails}
\label{sec:NTFPPorodTails}
The theory of phase-ordering kinetics builds on the definition of an order-parameter field $\psi(\vec{k},t)$ 
which, during coarsening, bears non-linear excitations such as solitary 
waves, domain walls, and defects.
A universal scaling form for the structure
factor $S(\vec k,t)=\langle\psi(\vec{k},t)\psi(-\vec{k},t)\rangle$ of the defect-bearing 
order-parameter field can be obtained by means of power counting. 

One generically considers an order parameter with $\mathcal{N}$ real-valued components, described by 
an $O(\mathcal{N})$-symmetric model Hamiltonian.
For such a field carrying a defect ensemble with a mean defect distance $\ell_{\text{d}}$, Porod's
law~\cite{poroda} (see \cite{PhysRevE.47.R9,Bray1994a} for a generalization to
arbitrary $\mathcal{N}$) predicts a scaling form for the angle-averaged
structure factor, in $d\geq\mathcal{N}$ spatial dimensions: 
\begin{equation}
  \label{eq:89}
  \int{\mathrm{d}\Omega_{\vec k}}\,S(\vec k,t) \sim \ell_\text{d}^{\,d}/(\ell_\text{d}k)^{d+\mathcal{N}}\,.
\end{equation} 
Considering a momentum range between the inverse defect distance and
an ultra-violet cut-off set by the size of the defect core, $1/\ell_\text{d}
\ll k \ll 1/\xi_h$, \Eq{89} follows from the $d$-dimensional Fourier
transform of a $d-\mathcal{N}$ dimensional but otherwise
structure-less defect in the order-parameter field.

For example, vortices in a 2D Bose condensate are zero-dimensional defects in the order-parameter
field $\phi$, \ie{}, we have $d=\mathcal{N}=2$. The scaling
\eq{58} of the corresponding bosonic occupation spectrum, for $\eta=0$, is equivalent to the
scaling form \eq{89}.
Hence, the prediction $\zeta=d+2-\eta$ for the scaling at a non-thermal fixed point 
 \cite{Berges:2008wm,Berges:2008sr,Scheppach:2009wu}, \cf{}~\Eq{58}, is consistent with the Porod law \eq{89}
 with $\mathcal{N}=2$, $\eta=0$.
Another example are magnetic domains in an order-parameter field obeying a $O(1)$ ($Z_{2}$) symmetric Hamiltonian such as for the Ising model.
The corresponding Porod exponent $\zeta=d+1$ also applies to domain walls such as solitons in a $d$-dimensional Bose gas and characterizes the spatial scaling of sound-wave turbulence, see, \eg{}~\cite{Khlebnikov2002a,Nowak:2011sk,Schmidt:2012kw}.

The rescaled momentum distributions at the non-thermal fixed points studied in the previous section can be fitted, in the infrared momentum region $k < k_\Lambda \simeq 0.5\,\xi_h^{-1}$, to a generalized
Cauchy distribution,
\begin{align}
  \label{eq:nkitofskappa}
  n(k,t_{0})=f_{s}\big(k/k_{\lambda}(t_{0})\big)\,,
  \\
  \label{eq:4}
  f_s(\kappa) = \frac{A}{1+\kappa^{\,\zeta}}\,,
\end{align}
shown as black lines in Figs.~\fig{vorlat_spectrum} and \fig{ranvort_spectrum}. Note that $A$ is a non-universal parameter which can be fixed by requiring a certain normalisation for the universal scaling function.
Thus, the collapsed
curve follows a power law $n(k,t_{0}) \sim k^{-\zeta}$ for $k_{\lambda}(t_{0}) \ll k \ll
k_\Lambda(t_{0})$. 
We find the Porod-tail exponents $\zeta=\zeta_{\text{g,a}}$ (see \App{fitting}),
\begin{align}
  \label{eq:zetagaussian}
  \zeta_{\text g}&\simeq 4.0\pm0.1\,,
  \\
  \label{eq:83}
  \zeta_{\text{a}} &= 5.7\pm0.3\,.
\end{align}
The scaling exponent \eq{zetagaussian} at the Gaussian fixed point ($\eta_{\text g}\simeq0$) is considerably smaller than $\zeta_{\text a}$, and is consistent with \Eq{89} for $\mathcal{N}=2$. 
The Porod tail with $\zeta_{\text g}=4$ results for dilute ensembles of randomly distributed vortices in $d=2$ dimensions, \cf{}~also Refs.~\cite{Nowak:2011sk,Schole:2012kt}.

For both non-thermal fixed points, the exponent $\zeta$ can be understood in terms of an extended Porod law.
To this end, we argue that, in general, the Porod law receives an additional correction from the anomalous dimension $\eta$, giving
\begin{align}
  \label{eq:PorodNeta}
  \zeta=d+\mathcal{N}-\eta.
\end{align}
The anomalous Porod exponent \eq{83} is consistent with  domains in the order-parameter field of an $O(1)$ ($Z_{2}$) symmetric model, \ie{}, $\mathcal N=1$, taking into account the independently determined anomalous exponent $\eta_{\text a}$. 
Inserting $\eta_{\text a}$ from \Eq{etaanomalous}, we obtain $\zeta_{\text a}=d+\mathcal{N}-\eta_{\text a}\simeq6$ consistent with our numerical result \eq{83}.

From our simulations, we infer that these domains are distinguished by the two possible orientations of the circulation (vorticity) a cluster of like-sign vortices can take.
The conjecture that the relevant order-parameter field is given by the vorticity density  $\omega\sim\nabla\times\vec{v}$ is further supported by the spectral decomposition of the momentum distribution shown in  \App{CompVsIncompSpectra}.
In \Fig{CompVsIncompSpectra}, the red triangles represent the contribution of the divergence-free part of the velocity field to the occupation-number spectrum $n(k)$ shown as black circles.
This part corresponds to the vorticity-bearing rotational flow caused mainly by the vortex defects, and its proximity to $n(k)$ demonstrates that it dominates the total spectrum in the region of the Porod-law fall-off, near both, the anomalous and the Gaussian fixed points \cite{Nowak:2011sk}.
This demonstrates that the vorticity plays an important role in the observed dynamics. 

We note that the angle-averaged vorticity spectrum scales as $\int\mathrm{d}\Omega_{k}\langle|\omega(\mathbf{k})|^{2}\rangle\sim k^{4}n(k)$ relative to the single-particle momentum distribution, cf.~Eq.~(37) of Ref.~\cite{Nowak:2011sk}.
Hence, supposing that vorticity represents the order parameter the coarsening of which can be described in terms of a $Z_{2}$-symmetric Landau-Ginzburg type model, one would expect, from our numerical result for $\zeta_\mathrm{a}$, \Eq{83}, the respective structure factor to exhibit a Porod tail $\sim k^{-2}$.
Assuming furthermore the validity of \Eq{PorodNeta} for this tail, within a so far not further specified description of phase-ordering kinetics, in $d=2$ dimensions and for $\mathcal{N}=1$, corresponding to the $Z_{2}$ symmetry, one obtains $\eta\simeq1$. 
A further examination of this is beyond the scope of the present work and will be done elsewhere.

We add the remark that a $Z_{2}$-symmetry breaking clustering transition has been described within an equilibrium formulation in Refs.~\cite{Yu2016a.PhysRevA.94.023602,Salman2016a.PhysRevA.94.043642}, with vorticity density as order parameter,  see also Ref.~\cite{Reeves2017a.arXiv170204445R} and our discussion in \Sect{DrivenStationarySystems} below.

In this context, it is also interesting that a scaling of $n$ with $\zeta=6$ could be associated with the scaling of the angle-averaged radial kinetic energy distribution $E(k) \sim k^3n(k) \sim k^{-3}$ corresponding to a
direct enstrophy cascade \cite{Schole:2012kt,Mininni2013a.PhysRevE.87.033002,Billam2014a.PhysRevLett.112.145301,Billam2015a.PhysRevA.91.023615,Reeves2017a.arXiv170204445R} known to characterize classical turbulence in a continuously driven 2D incompressible fluid \cite{Kraichnan1967a}.
This power law is steeper than that of a classical Kolmogorov-$5/3$ law $E(k) \sim k^{-5/3}$, corresponding to
$\zeta=4.66$.

Note, furthermore, that, as seen in \Fig{vorlat_spectrum_unresc}, the power law \eq{83} is found already at very early times, when the vortices still form strong clusters.
We recall that an elementary vortex, on scales smaller than the size of its core, gives rise to a power-law momentum spectrum $n(k)\sim k^{-6}$, \cf{}~Fig.~7 of Ref.~\cite{Nowak:2011sk}.
This scaling is here found to survive after the break-up of a non-elementary vortex into elementary vortices, over scales on the order of the cluster size.

We finally note that the $\eta$ term in \Eq{PorodNeta} is also corroborated by the single-particle spectra of \emph{weak sound-wave turbulence} computed numerically in Refs.~\cite{Khlebnikov2002a,Nowak:2011sk}.
In this case, $\mathcal N=1$ should also apply because the turbulent transport of compressible (sound-wave) excitations is expected to involve solitary waves which, as mentioned above, represent defects of an $O(1)$-symmetry breaking order-parameter field.

As was furthermore demonstrated in Ref.~\cite{Mathey:2014xxa}, numerically determined spectra of compressible excitations for $d=1,2,3$ are consistent with scaling exponents $\zeta$ inferred from renormalization-group analyses of the Kardar-Parisi-Zhang (KPZ) equation.
Both, the numerically obtained exponents $\zeta$ for the compressible excitations, and the exponents determined in the KPZ framework, can be explained by the same non-zero anomalous dimension $\eta$.
This provides a further support of the anomalous correction in \Eq{PorodNeta}.

%======================================================================================
\subsection{Coarsening as a form of phase-ordering kinetics}
Arguments based on classical dynamics~\footnote{The
  calculations assume $O(\mathcal{N})$ models within the Hohenberg--Halperin
  classification~\cite{Hohenberg1977a}} 
of the order-parameter field~\cite{PhysRevE.47.R9}, or on a renormalisation group
approach~\cite{PhysRevLett.62.2841,Bray1994a}, allow
to predict the temporal scaling of the defect length as
\begin{equation}
  \label{eq:5}
  \ell_\text{d}(t)\sim t^{\,\beta_{\text d}}\,,   
\end{equation}
neglecting possible multiplicative logarithmic corrections. 
The exponent $\beta_{\text d}$ characterizes
the speed of the coarsening process, which is influenced by
conservation laws in the order-parameter field.

The coarsening dynamics of the order-parameter field is typically described by a diffusion-type equation \cite{Bray1994a},
\begin{equation}
  \label{eq:90}
  \left(\frac{1}{ak^\mu} +\frac{1}{\Gamma}\right)\partial_t \psi(\vec{k},t) = \frac{\delta F[\psi]}{\delta \psi(-\vec{k},t)}\,,
\end{equation}
where  
$F[\psi]$ is the free energy of the order-parameter field $\psi$
defining the model under consideration. 
The term involving the diffusion constant $\Gamma$ applies in the case of 
non-conserved fields, while the first term on the left-hand side
accounts for additional conservation laws the field obeys, parametrized by a transport coefficient $a$ \cite{Bray1994a}.
The parameter $\mu$ defines the nature of the conservation law which effectively modifies the
scaling of the kinetic (typically Laplacian) operator in the equation of motion.

For example, coarsening dynamics near equilibrium, according to the 
purely dissipative model A within the Hohenberg--Halperin classification 
\cite{Hohenberg1977a}, has  $\mu=0$, while $\mu=2$ applies to the diffusive model B for conserved order parameters.

The coarsening process described by \Eq{90} is governed by the scaling exponent, 
\begin{equation}
  \label{eq:BraybetaNgtT}
       \beta_{\text d} = \frac{1}{2+\mu}\,,
\end{equation}
if $\mathcal{N} > 2$ or $\mu=0$, while for $\mathcal{N} < 2$ and $\mathcal{N}+\mu>2$ one finds
\begin{equation}
  \label{eq:BraybetaNltT}
       \beta_{\text d} = \frac{1}{\mathcal{N}+\mu}\,.
\end{equation}
In the marginal case $\mathcal{N} =2$ and $\mu>0$, a logarithmic correction applies, 
\begin{equation}
  \label{eq:ldtforN2}
  \ell_\text{d}(t)\sim (t\ln t)^{1/(2+\mu)}\,, 
\end{equation}
\cf{}~Ref.~\cite{Bray1994a} and in particular Fig.~$24$ therein for other cases with $\mathcal{N}+\mu\leq2$.

Inserting the scaling \eq{5} into the spatial scaling form \eq{89},
neglecting potential logarithmic corrections,
and including a simple infrared cutoff in terms of a constant $C$
results in a scaling form for the structure factor,
\begin{equation}
  \label{eq:BrayScaling}
  S(\vec k,t) \sim \frac{t^{\,d\,\beta_{\text d}}}{1+C\,(t^{\,\beta_{\text d}}k)^{\,d+\mathcal{N}}}\,.
\end{equation}
Assuming that the scaling form \eq{BrayScaling} and the scaling relation \eq{NTFPScaling} 
describe the same physical processes, one obtains the scaling relations $\beta = \beta_{\text d}$
and $\alpha = d\beta_{\text d}$ between the scaling exponents \footnote{This argument does not exclude the possibility of a different scaling behavior in the infrared limit, $k\to0$, as, \eg{}, found numerically in Ref.~\cite{Orioli:2015dxa}.}.

This suggests that the defect coarsening, up to logarithmic corrections, is in one-to-one correspondence
with the self-similar transport process near the non-thermal fixed point. 
For $\beta_{\text{d}}>0$, the transport is directed towards infrared momenta and is subject to particle conservation
in the scaling regime (see Eqs.~\eq{2} and \eq{ParticleConserv}).

%======================================================================================
\subsection{Coarsening at the non-thermal fixed points}
\label{sec:CoarseningAtNTFPs}
%
%==========================================
\begin{figure}[!t]
\centering
    \includegraphics{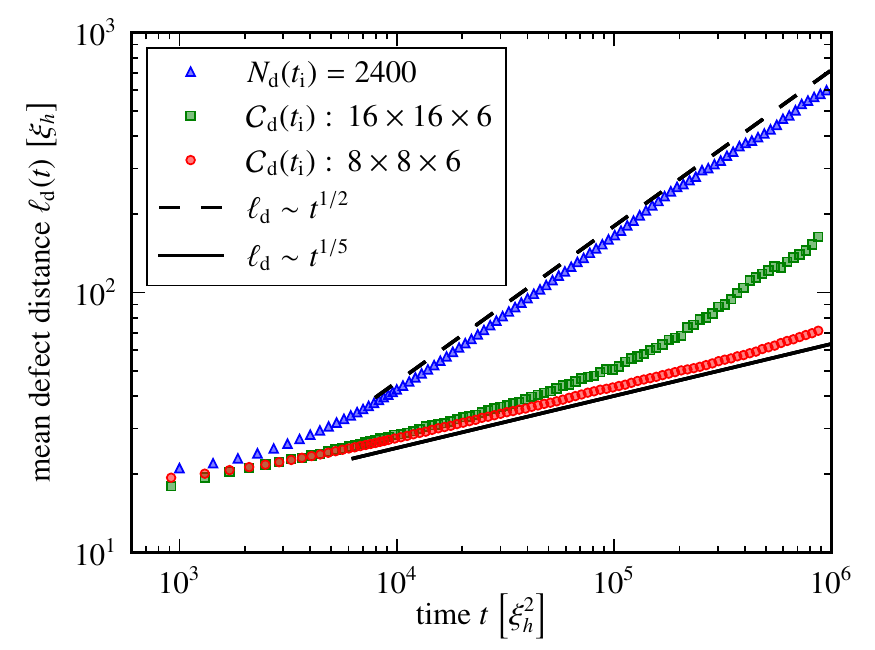}
    \caption{Mean defect distance $\ell_\text{d}$ as function of time,
      after phase-imprinting different initial vortex
      configurations at $t_{\text i}=0$. 
      The (blue) triangles show the evolution from a random distribution of
      $N_{\text{d}}(t_{\text i}) = 2400$ elementary vortices and anti-vortices
      (total angular momentum zero), corresponding to the evolution
      shown in \Fig{ranvort_spectrum}.
      The (green) squares and (red) circles show the evolution from 
      an irregular square lattice of $16\times16$ and $8\times8$ non-elementary
      vortices with winding numbers $w=\pm6$, see \Fig{vorlat}a for an example.
      The evolution marked by the (red) circles corresponds to \Fig{vorlat_spectrum}.
      As indicated by the dashed and solid black lines, the growth of the distance
      is well approximated by power laws $\ell_\text{d}(t)\sim t^{\,\beta_{\text d}}$ 
      with  $\beta_{\text d}={1/2}$ (Gaussian fixed point) and $\beta_{\text d}={1/5}$ 
      (anomalous fixed point), respectively. 
      As indicated by the evolution marked by the (green) squares, depending on the 
      initial conditions, the system can
      first approach the anomalous fixed point before it turns to steeper scaling
      near the Gaussian one. 
     }
  \label{fig:thermvsiso}
\end{figure}
%==========================================

With the above results from the theory of phase-ordering kinetics at hand, 
we can analyse the dynamics near the previously described non-thermal fixed points
by evaluating the time evolution of the average defect distance $\ell_{\text{d}}(t)$.
\Fig{thermvsiso} shows $\ell_{\text{d}}(t)$ for the different initial vortex configurations chosen in our numerics. 
Blue triangles mark the evolution from the 
random initial distribution of $N_\text{d}(t_{\text i}) = 2400$ elementary vortices and anti-vortices, 
for which the occupation spectrum is shown in \Fig{ranvort_spectrum} and 
a scaling evolution with exponents \eq{alphabetag} is found at late times, indicating the approach of the (near-)Gaussian fixed point.

The green squares and red circles show the evolution from a vortex-lattice initial state, containing
defects with winding number $w = \pm 6$.
The green data starts from a lattice of $16\times 16$ vortices, the red data from an $8 \times 8$-lattice.
The evolution of the occupation spectrum corresponding to the red data
is shown in \Fig{vorlat_spectrum} and gives an anomalously slow scaling evolution, indicating the approach to the
anomalous non-thermal fixed point \eq{82}. 

\Fig{thermvsiso} demonstrates that, for both types of initial conditions chosen
here, the evolution of the length scales is consistent with scaling
behaviour, $\ell_{\text{d}}(t) \sim t^{\,\beta_{\text d}}$, within certain intervals of time.
For example, for both, the blue and the red data, scaling is seen for $t>t_0\simeq
10^4\,\xi_h^2$, simultaneously with the scaling evolution of the occupation spectrum.

The evolution marked by the green squares in \Fig{thermvsiso} involves an initial decay of the non-elementary 
vortices into $1536$ elementary defects.
At later times, the growth of  $\ell_{\text{d}}(t)$ (green squares) is also found to approach the Gaussian fixed point.
This shows that the system can switch dynamically
between the two different types of phase-ordering kinetics. As each
type is associated with a different non-thermal fixed point, this
demonstrates that the system first approaches
the anomalous fixed point and only later is attracted by the Gaussian one.

We note that whether and how long the anomalous fixed point is approached appears to be mainly determined by the initial density of defects rather than by their spatial arrangement in the system.
Starting with winding numbers larger than one seems to facilitate the approach to the anomalous fixed point. 
A detailed study of the different conditions determining the attractive basin of the anomalous fixed point is beyond the scope of the present work.
As we will show, however, in the next subsection, the effective coupling between the defects and the fluctuating bulk controls the transition from the strongly anomalous to the (near-)Gaussian scaling behavior.

Near both non-thermal fixed points the coarsening exponents are found to be consistent with each other. 
Near the Gaussian fixed point,  $\beta_{\text d} \simeq 0.5 \simeq \beta_{\text{g}}$ and 
$\alpha_{\text{g}} \simeq d\beta_{\text d} \simeq 1$ (\cf{}~\Eq{alphabetag}). 
These values are consistent with 
Bray's prediction, \Eq{BraybetaNgtT}, if purely dissipative dynamics,
\begin{equation}
  \label{eq:mugaussian}
  \mu_{\text g} \simeq 0\,,
\end{equation}
is assumed for the order-parameter field. 

The value $\beta_{\text a} = 0.193\pm0.05$ we find near the anomalous  fixed point
(\cf{}~\Fig{vorlat_spectrum} and \Eq{82}), characterizes a distinctly different class of phase-ordering kinetics.
Taking into account that $\mathcal{N}=1$, as argued in \Sect{NTFPPorodTails}, and setting $\beta_{\text d}=(\mathcal{N}+\mu)^{-1}=\beta_{\text a}=(2-\eta_{\text a})^{-1}$, \cf{}~Eqs.~\eq{BraybetaNltT} and \eq{2}, respectively, the anomalous exponent 
$\eta_{\text a}\simeq -3$ maps to 
\begin{equation}
  \label{eq:muanomalous}
  \mu_{\text a} \simeq 4\,.
\end{equation}

In the theory of phase ordering kinetics, $\mu>0$ 
signals that the time-evolving order parameter fulfills 
additional conservation laws \cite{Bray1994a}.  
Formally, in the language of renormalization-group theory, $\mu$ takes the 
role of the scaling exponent $\eta$ of a wave-function
renormalization factor.
This corroborates our conjecture, \Eq{muanomalous}.
Again, we have neglected any possible logarithmic corrections. 

We remark that, however, the prediction \eq{2} for $\beta$ results from a next-to-leading-order large-$\mathcal N$ expansion \cite{Orioli:2015dxa}. 
Hence, for large $\mathcal N$ where it is expected to be valid rigorously, the relation \eq{BraybetaNgtT} applies, and thus, for $\mathcal{N}\ge2$,  
\begin{equation}
  \label{eq:muvsetaNg2}
  \mu = -\eta\,.
\end{equation}
This argument also provides a possible explanation for the independence of the scaling exponent $\beta$ of $\mathcal{N}$ \cite{Orioli:2015dxa}.
The energy of the order parameter field per unit defect-core volume is, for $\mathcal{N}\ge2$, mainly given by the contribution from the field around the defect core the scaling of which is fixed by geometry \cite{Bray1994a}.
For example, the velocity field around point defects in $d=2$ dimensions scales as $1/r$ in the distance $r$ from the core.
Hence, if coarsening is dominated by (radially symmetric) defects, this contribution fixes $\beta=1/2$, for $\mathcal N\ge2$, up to the anomalous correction $\eta$.

We furthermore note that our results are also compatible with the scaling observed for the scaling dynamics of relativistic $O(\mathcal{N})$ models, for $\mathcal{N}=2$ and $4$, giving $\beta\simeq0.5$ \cite{Orioli:2015dxa}. 
As demonstrated in Ref.~\cite{Gasenzer:2011by} for the case $\mathcal{N}=2$, charge domains separated by vortex sheets characterize the evolution near the non-thermal fixed point, consistent with the conditions for a coarsening exponent $\beta=1/(2-\eta)$, with $\eta\simeq0$. 
We finally remark that scaling which reflects the occurrence of more intricate topological defects beyond vortices \cite{Moore:2015adu}, possible in $O(\mathcal{N})$ models for $\mathcal{N}>2$, is expected to require a higher spatial dimensionality, $d\geq3$.

In summary,  the universal scaling corresponding to the coarsening at the Gaussian non-thermal fixed point is consistent with the class defined by the Hohenberg--Halperin model A. 
We emphasize that, however, the dynamics at the far-from-equilibrium non-thermal fixed point is not necessarily equivalent to near-equilibrium coarsening at the fixed point describing the corresponding equilibrium phase transition \cite{Bray1994a}.
The coarsening related to the anomalous non-thermal fixed point does not fall, to our knowledge, in any of the known standard classes of dynamical critical phenomena.

%======================================================================================
\subsection{Coarsening under coupling to a thermal bath}
%
%==========================================
\begin{figure}[!t]
\centering
    \includegraphics{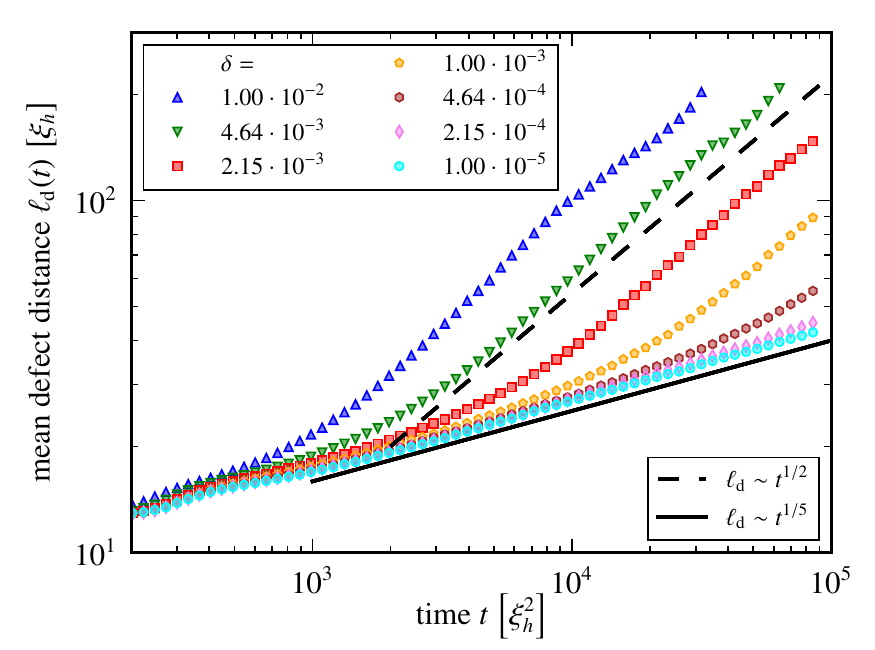}
    \caption{Mean defect distance $\ell_\text{d}$ as function of time,
      after preparing an irregular $8\times8$ square lattice of non-elementary 
      (anti-)vortices with winding number $w=\pm6$.
      Symbols (colors) mark the evolution for different parameters $\delta$ and $\gamma$, 
      describing the coupling to the bath, \cf{}~\Eq{Titoalphagamma}, in a way that the temperature is
      held fixed at $T=2k_\text{B}\delta/\gamma=100\,\xi_h^{-2}$. 
      Comparing with \Fig{thermvsiso} we find  the same universal 
      growth laws $\ell_\text{d}(t)\sim t^{\,\beta_{\text d}}$ near the anomalous fixed point 
      ($\beta_{\text d}={1/5}$) and near the Gaussian fixed point ($\beta_{\text d}={1/2}$) 
      as for the isolated system.
      Moreover, the crossover time scale from anomalous to Gaussian scaling is 
      controlled by the strength of the coupling $\delta$ to the thermal bath.
      This suggests that also through the 
      different initial conditions chosen in \Fig{thermvsiso}, a thermal bath is created
      which eventually forces the system to deviate from the anomalous fixed point.
      At late times, the data is limited by averaging statistics while at least one vortex--anti-vortex pair is left.
      }
  \label{fig:ThermalCoarsening}
\end{figure}
%==========================================
Our results discussed above suggest that the
coarsening dynamics near the Gaussian non-thermal fixed point in the isolated 2D Bose gas
is consistent with coarsening.
This coarsening occurs according to the Hohenberg--Halperin model A, \ie{}, within a classical dissipative setting, following quenches across the corresponding thermal
phase transition.
To investigate this relation further, we present, in the following, analogous results
for coarsening in a dilute 2D Bose gas coupled to a thermal bath.

As before, we choose the same type of non-equilibrium initial conditions by setting a weakly 
irregular rectangular lattice of non-elementary vortices.
At time $t_{\text i}$, we couple the system to a thermal bath and compute the ensuing
time evolution according to the driven-dissipative stochastic Gross-Pitaevskii equation 
\cite{Cockburn2012a}.
The temperature is chosen such that, at late times, the system approaches a thermal state
without free vortex defects below the Berezinskii--Kosterlitz--Thouless (BKT) transition.
Details of the numerical method and results showing the self-similar coarsening evolution of the occupation spectrum 
are given in \App{ThermalCoarsening}.

\Fig{ThermalCoarsening} shows the scaling found in the time evolution of the mean 
defect distance $\ell_{\text d}$, in analogy
to \Fig{thermvsiso} for the isolated case.
As before, we start from an irregular $8\times8$ square lattice of non-elementary (anti-)vortices with winding number $|w|=6$.
We vary the parameters $\delta$ and $\gamma$, describing the coupling to the bath, \cf{}~\Eq{Titoalphagamma}, in a way that the temperature is held fixed at $T=100\,\xi_h^{-2}$.
\Fig{ThermalCoarsening} shows the ensuing time evolution of $\ell_{\text d}(t)$ for a range of different $\delta$ as indicated.
For stronger coupling, Gaussian scaling evolution with $\ell_{\text d}(t)\sim t^{1/2}$ is seen during the late-time evolution.
As before, in addition to a pure scaling behavior, our numerical results are in principle consistent with weak logarithmic corrections \cite{Bray1994a,Bray2000PhRvL..84.1503B}.

Reducing, however, the coupling $\delta$ to the bath, the initial vortex clustering is found to survive for a sufficiently long time such that we are again observing anomalous non-thermal-fixed-point scaling at early times of the evolution, before the system switches to Gaussian scaling.
The cross-over to thermal scaling occurs the earlier, the higher the coupling is, at a time $t_{\text c}$ which approximately scales as $t_{\text c}\sim\delta^{-1.2\pm0.2}$.

Hence, as before, our results indicate that the chosen initial vortex ensemble induces the system 
to be driven closely to the anomalous fixed point at early times of the evolution.
We furthermore infer that the late-time Gaussian scaling observed in the isolated system can be
associated with a thermal bath built up self-consistently in the system.
This implies that the macroscopic parameter controlling the type of fixed point behavior in the isolated system is the effective dissipation to the fluctuating bulk, which builds up depending on the energy in the UV induced through the initial vortex distribution.
This leads us to the conjecture that the Gaussian non-thermal fixed point in the isolated system
is consistent with the behavior of the coarsening dynamics predicted within the theory of phase-ordering
kinetics.

%==========================================
\begin{figure}[!t]
\centering
    \includegraphics{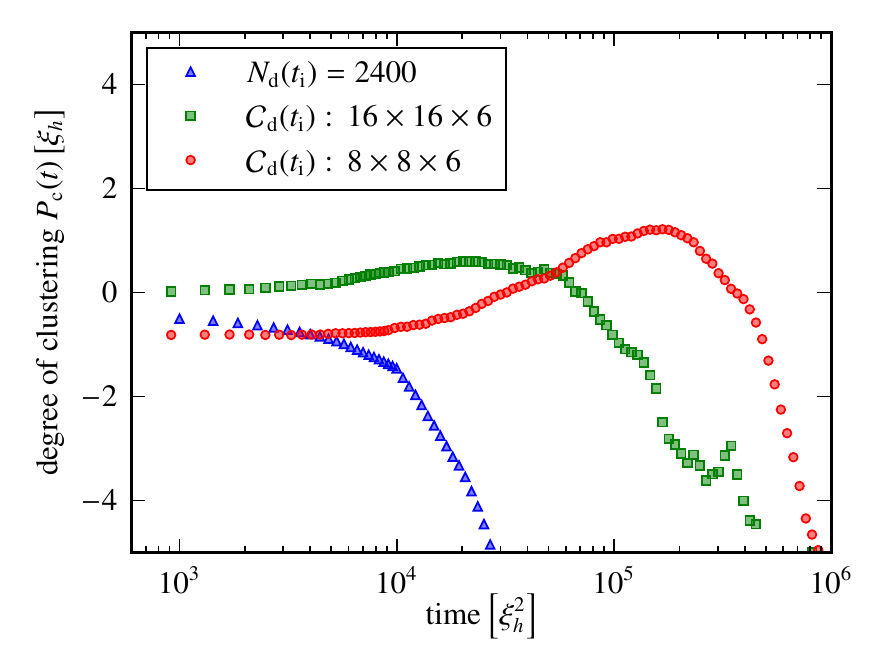}
    \caption{%
    Degree $P_{\text c}(t)$ of clustering of elementary like-sign vortices as a function of time,  
    defined in \Eq{PClustering}, for the same three initial conditions as used in \Fig{thermvsiso}. 
    Deviations of $P_{\text c}$ from zero indicate clustering ($P_{\text c}>0$) or anti-clustering ($P_{\text c}<0$). 
    A strong anti-clustering of like-sign defects, such as for the data marked by (blue) triangles, 
    is accompanied by vortex--anti-vortex pairing.
    }
  \label{fig:cluster}
\end{figure}
%==========================================

%======================================================================================
\subsection{Three-body-collision induced vortex loss}
\label{sec:ThreeBodyKinetics}
The anomalously slow coarsening behavior can be understood, in leading asymptotic approximation, in terms of the scattering kinetics of vortex defects created in the system.
The essential characteristics of this kinetics is seen rather clearly in a motional visualization of the dynamics 
\footnoteremember{fn:Video}{Videos of exemplary runs of the coarsening kinetics can be found under \url{http://www.kip.uni-heidelberg.de/gasenzer/projects/anomalousntfp}.}.
As discussed above, the initial states containing non-elementary vortices are seen to exhibit a particularly strong clustering of like-sign defects during the early stage.
This observation is corroborated by the results shown in \Fig{cluster} for the degree $P_{\text c}(t)$ of clustering at time $t$.
This measure is defined as
\begin{equation}
  \label{eq:PClustering}
  P_{\text c}(t) = \int_a^b\!\text{d}\ell\, \left[\mathcal{L}(\ell,t)-\ell\right]/(b-a)\,,
\end{equation}
in terms of Ripley's $\mathcal{L}$ function,
\begin{equation}
  \label{eq:RipleyL}
  \mathcal{L}(\ell,t) = \left\{V\left[\pi N_{\text v}(t)^{2}\right]^{-1}\sum_{i\not=j}\theta\left(\ell-d_{ij}(t)\right)\right\}^{1/2}\,,
\end{equation}
where $V$ is the system's volume, $N_{\text v}(t)=N_{\text d}(t)/2$ is the number of like-sign defects at time $t$.
The $\theta$ function counts all pairs of like-sign defects $i,j$ with Euclidean distance $d_{ij}$ below a maximum of $\ell$.
We chose the integration interval to be bound by $a=6\,\xi_h$ and $b=150\,\xi_h$.
Thus, positive values of $P_{\text c}$ indicate clustering while for $P_{\text c}<0$ strong anti-clustering is present, \ie{}, pairing of vortices and anti-vortices.
Comparing with \Fig{thermvsiso}, clustering of like-sign defects appears to be correlated with the anomalously slow coarsening.

Examining the motional visualization \footnoterecall{fn:Video} in more detail one observes that, during the anomalous coarsening, the defects are rather evenly distributed across the volume.
While pairs and clusters of vortices with same-sign circulation rotate around each other, pairs of loosely bound vortices and anti-vortices (vortex `dipoles') travel coherently along the boundaries of the clusters, dragged along by the strong flux.
Thereby, loss of vortices appears to be caused by three-body collisions of defects with pairs of both, even-sign and opposite-sign circulation. 
These collisions allow vortices with opposite circulation to approach each other more closely than a few times the healing length $\xi_\mathrm{h}$.
As a result, these vortices can form closely bound pairs moving at a high speed, which are less tied to the large-scale coherent flow. 
Soon after their formation, these pairs can mutually annihilate by interaction with the background noise.

The motion of point vortices can be described by the Onsager model \cite{Onsager1949b}.
This model accounts for an attractive interaction between vortices and anti-vortices which scales with the logarithm of the defect distance. 
However, the Hamiltonian does not possess a kinetic term which would allow this potential energy to be transferred to kinetic energy and thus enable a direct mutual approach of the defects.

Only at short distances, below a certain threshold of a few healing lengths $\xi_\mathrm{h}$, additional dissipative forces act on the vortices due to the background noise.
These forces are not accounted for by the basic Onsager model and allow pairs of vortices to do the final move, viz.~directly approach each other and mutually annihilate. 
During this process, they are still moving at high speed, as a Helmholtz pair, along the direction perpendicular to the dipole-vector.

At larger distances, the only way for a vortex to get close to an opposite-sign vortex and for a dipole to further reduce its binding length, is to scatter with a another vortex or dipole whereby energy can be transferred.
As the Onsager potential energy scales logarithmically with distance, the defects taking part in the collision should have similar distances.
For recent theoretical studies of such processes, see Refs.~\cite{Schole:2012kt,Simula2014a.PhysRevLett.113.165302,Groszek2016a.PhysRevA.93.043614}.

A simple kinetic argument supports our conjecture that, near the anomalous fixed point, three-body scattering prevails while  two-body collisions dominate near the Gaussian fixed point:
Suppose a kinetic equation  $\partial_t N_{\text d} \sim -\Gamma N_\text{d}$ accounts for the mean decrease of the number of defects $N_{\text d}(t)$, with the collision rate $\Gamma$ depending on $N_\text{d}$ itself.

The two-body collision rate of vortices with mean free traveling time $\tau$ can be estimated to scale as $\Gamma_{2}\sim\tau^{-1}$.
That is, within the time
\begin{equation}
  \label{eq:tau}
  \tau \sim \frac{l_\mathrm{mfp}}{v}\,,\qquad\mbox{where}\qquad
  l_\mathrm{mfp} \sim \frac{V}{N_\text{d}\sigma}\,,
\end{equation}
on average, a given vortex comes close to another one.
Here, $l_\mathrm{mfp}$ is the mean free path of the vortices, with cross section a few times the healing length and thus scaling as $\sigma\sim\xi_\mathrm{h}$, and $v$ is the mean velocity of the vortex.

Consider now the probability that, at the moment of the collision, a further vortex is closer than a small multiple of the healing length to one of the scattering partners.
Using our above results, we estimate this probability to scale as $\xi_\mathrm{h}/l_\mathrm{mfp}\sim\xi_\mathrm{h}^{2}N_{\text d}/V$.
This is of the order of the fraction of the mean volume per vortex, $V/N_{\text d}$, which is occupied by a closely bound pair with a linear extent of a few times $\xi_\mathrm{h}$.

As a result, only a fraction $\sim\xi_\mathrm{h}^{2}N_{\text d}/V$ of all vortex-vortex encounters can be counted as three-body collisions.
The resulting rate of vortex loss through three-body scattering is therefore estimated to scale as the product of $\Gamma_{2}$ and this fraction. 
Hence the rate is scaling as
\begin{equation}
\label{eq:gamma3}
-(N_{\text d}^{-1}\partial_{t}N_{\text d})
\sim\Gamma_{3} 
\sim \Gamma_{2}\,\xi_\mathrm{h}/l_\mathrm{mfp}
\sim\tau^{-1}N_{\text d}\,\xi^{2}_\mathrm{h}/V\,.
\end{equation}
Inserting $\tau$ from \Eq{tau} one obtains the rate to scale as $\Gamma_{3}\sim vN_{\text d}^{2}\,\xi^{3}_\mathrm{h}/V^{2}$.
Note that $\Gamma_{3}\sim N_{\text d}^{2}$ is the standard scaling of a three-body scattering rate between scatterers with concentration scaling as $N_\mathrm{d}$.

The crucial difference here is that the velocity $v$ varies with $N_\mathrm{d}$ as well.
The rather uniform distribution of vortices and anti-vortices within the system means that their mean velocity scales with the inverse of the mean pair distance, $v\sim1/l_{\text D}$, which is on the order of the mean defect distance, $l_{\text D}\sim\ell_{\text d}\sim(V/N_{\text d})^{1/2}$. 
As a result, we find
\begin{equation}
\label{eq:ThreeBodyKinetic}
   \partial_{t}N_{\text d}\sim-\Gamma_{3}N_{\text d} 
   \sim -\,\mbox{const.}\times N_{\text d}^{7/2}\,,
\end{equation}
where the constant has units of $t^{-1}$, having the scaling solution
\begin{equation}
  \label{eq:ThreeBodyDecay}
  N_{\text d}(t) \sim t^{-2/5}\,.
\end{equation}
This is consistent with the increase $\ell_{\text d}(t)\sim[V/N_{\text d}(t)]^{1/2}\sim t^{1/5}$ of the defect spacing seen in Figs.~\fig{thermvsiso} and \fig{ThermalCoarsening}.

The above kinetics is different from that near the Gaussian fixed point.
This can be inferred  from the motional visualization  \footnoterecall{fn:Video} which shows that vortices are predominantly already paired with each other, with different pairing distance. 
Clustering of like-sign vortices and concomitant  large-scale flows do not appear. 
As a consequence, dipoles do not have to be formed first, and mutual annihilation follows quickly after two-body scattering of a fast-moving vortex pair with another defect.
This is described by a kinetic equation  
\begin{equation}
  \label{eq:TwoBodyKinetic}
  {\partial_{t} N_{\text d}} \sim - \Gamma_{\text D}N_{\text d}\,,
\end{equation}
where the rate scales with the mean free traveling time of pairs, $\Gamma_{\text D}\sim\tau_{\text D}^{-1}\sim N_{\text d}\sigma_{\text D} v/V$, with cross section $\sigma_{\text D}\sim\xi_{\text h}$, and velocity $v\sim\ell_{\text D}^{-1}\sim\xi_{\text h}^{-1}$.
Hence $\Gamma_{\text D}\sim N_{\text d}$ and thus $N_{\text d}(t)\sim t^{-1}$, consistent with $\ell_{\text d}(t)\sim t^{1/2}$.
This is  seen in our data, see also previous work in Ref.~\cite{Schole:2012kt}.

We conclude that the anomalously slow coarsening process is dominated by the direct interactions between three defects, not obstructed by dissipative interactions with the bath of thermal or near-thermal small-scale fluctuations \cite{Damle1996a.PhysRevA.54.5037}. 
In contrast, the main decay mechanism of vortices in the thermally coupled gas is provided by vortex--anti-vortex annihilation after a comparatively fast diffusive mutual  approach of the defect pair.
This is due to the fact that the drift under the Magnus force, in the dissipative system, exceeds the parallel propagation of  pairs of opposite-sign defects according to the Helmholtz vortex law.

The dynamics of the vortex distributions at the anomalous fixed point is likely to have strong relations to 2D classical turbulence.  
Kinetic energy spectra and possible relations to classical turbulence were studied in
Refs.~\cite{Bradley2012a,Reeves2013a.PhysRevLett.110.104501,PhysRevLett.110.104501,Billam2014a.PhysRevLett.112.145301,Yu2016a.PhysRevA.94.023602,Salman2016a.PhysRevA.94.043642,Reeves2017a.arXiv170204445R}
exhibiting the relevance of the formation of like-sign vortex clusters for the energy spectra to show Kraichnan-type scaling behaviour. 
In contrast to this, starting from different initial conditions, the defects tend to arrange in randomly distributed bound vortex--anti-vortex pairs showing, at early times, two-body scattering loss, before being slowed down to a three-body-dominated decay later~\cite{Schole:2012kt}.

%======================================================================================
\subsection{Relation to scaling behavior in driven stationary systems}
\label{sec:DrivenStationarySystems}
Our findings motivate us to ask whether the vortex configurations can be stabilized in a way such that the dynamical scaling behaviour at the anomalous non-thermal fixed point can  be associated with an appropriate \emph{equilibrium} fixed point.
In the remainder of this chapter we briefly discuss possible interpretations in the context of near-equilibrium dynamical critical phenomena, in particular of vortex glasses in type-II superconductors and a vortex-clustering phase transition in a dilute 2D Bose gas.

The renormalisation-group approach to defect
coarsening~\cite{PhysRevLett.62.2841,Bray1994a} allows
for a more intuitive interpretation of the anomalously slow phase-ordering 
process. 
Generally, in making a scaling hypothesis for a correlation function, one needs
to introduce a dynamical critical exponent $z_{\text d}$, defining the relation
between the temporal and spatial scaling of the renormalization flow of the
order-parameter field~\footnote{$z_{\text d}$ applies to
  the scaling of the defect-bearing order-parameter field and does not necessarily
  equal the dynamical critical exponent $z$ of microscopic quasi-particle 
  excitations close to the non-thermal fixed point, \cf{}, \eg{}, the analysis
  in Ref.~\cite{Orioli:2015dxa}. 
  Note also that the dependence on $z$ drops out for the 
  particle-transport exponents $\alpha$ and $\beta$, \cf{}~\Eq{2}}.
For the structure factor, the scaling hypothesis reads
\begin{equation}
  \label{eq:SktScaling}
  S(s\vec k,s^{-z_{\text d}}t) = s^{-\alpha_{\text d}z_{\text d}}S(\vec k,t)\,.
\end{equation}
Comparing this relation with the scaling forms \eq{89} and \eq{BrayScaling}, 
we find that the temporal scaling of $\ell_\text{d}(t)\sim
t^{\,\beta_{\text d}} \sim t^{1/z_{\text d}}$ provides the dynamical exponent in terms of 
$\beta_{\text d}$,
\begin{equation}
  \label{eq:94}
  z_{\text d}=1/\beta_{\text d}\,, 
\end{equation}
see also Ref.~\cite{Schachner:2016frd}.
The dynamical exponent $z_{\text d}$ is associated with 
the particular dynamic universality class which the near-critical evolution of the  
effective, defect-bearing order-parameter field falls into. 
Hence, our results imply an anomalously
large value of the dynamical critical exponent, 
\begin{equation}
  \label{eq:zdanomalous}
  z_{\text d} \simeq 5\,. 
\end{equation}
In contrast, thermally diffusing vortices near the Gaussian fixed point
show $z_{\text d}=2$  \cite{Bray1994a}, and thermal quenches in the superfluid phase of the
two-dimensional Bose gas give rise to dynamics governed by 
$z_{\text d}=1$~\cite{Damle1996a.PhysRevA.54.5037}.

Considering driven stationary systems, a dynamical critical exponent $z=5$, determining the relative scaling in the frequency dependence of the dynamic structure factor and the spectral function, is, as in phase-ordering kinetics, most compatible with a conserved order parameter, such as in the Hohenberg--Halperin model B~\cite{Hohenberg1977a}. 
Experimental evidence for a value $z\approx 5$ has been reported near a transition from a vortex-lattice phase to a vortex-glass phase~\cite{PhysRevLett.63.1511, PhysRevLett.66.953}, studied extensively in the context of type-II superconductors \cite{Nattermann2000AdPhy49607N}.
The glass appears when a regular Abrikosov lattice changes into a phase of disordered, pinned vortices, below the transition to a vortex liquid of more freely propagating defects. 
The vortices affect the current-voltage characteristics of the superconductor in a way which depends on the density, position, and interactions of the defects. 

Numerical studies~\cite{PhysRevB.44.9780} corroborate the measured value of $z$, and theoretical models for a vortex-glass phase in superconductors have been proposed~\cite{PhysRevB.43.130,PhysRevB.45.523}.  
For a complex Ginzburg-Landau model with a static magnetic field and a Gaussian stochastic mass, van Hove scaling results, $z=2(2-\eta)$, corresponding to a conserved order parameter and an overdamped zero-frequency hydrodynamic mode.
A first-order epsilon expansion around the respective upper critical dimension, $d_{\text{up}}=6$, yields
$z=2(2+\epsilon/6)$ \cite{PhysRevB.45.523}, similarly as for Ising spin glasses \cite{Zippelius1984a.PhysRevB.29.2717}.
In $d=2$ dimensions, this extrapolates to an order-of-magnitude estimate for the dynamical exponent of $z\simeq5.3$.
In a charge-neutral dilute Bose gas, an Abrikosov lattice can be created by means of rotating-phase laser fields \cite{Abo-Shaeer2001a,Zwierlein2005a.Natur.435.1047Z} and could, possibly, open new perspectives to study dynamical critical behavior in such systems.

For charge-neutral atoms in a closed trap without external fields, a potentially related clustering transition has been reported \cite{Yu2016a.PhysRevA.94.023602,Salman2016a.PhysRevA.94.043642}.
Onsager's point-vortex model \cite{Onsager1949b} implies a maximum entropy state in which an equal number of elementary vortices and anti-vortices are randomly distributed in the system. 
It was shown that, continuously driven, such a system can exhibit a dynamical transition from a state described by a positive temperature to one with negative temperature, depending on whether the entropy rises or decreases with increasing energy density.
While, for positive temperatures, vortices and anti-vortices mingle and form bound pairs,  like-sign vortices are expected to spontaneously cluster with each other at negative temperatures.

Our results indicate that, during the coarsening process, the scaling exponent $\beta$ switches from the anomalous to the Gaussian value, when the clustering of like-sign defects disappears, see Figs.~\fig{thermvsiso} and \fig{cluster}.
Hence, it is imaginable that in a driven situation, with a vortex ensemble containing a constant number of defects on average and representing a state close to the clustering transition of Refs.~\cite{Yu2016a.PhysRevA.94.023602,Salman2016a.PhysRevA.94.043642}, the respective critical dynamics is related to the anomalous fixed point we report here.

%======================================================================================
%
\section{Conclusions}
\label{sec:conclusions}
In this work, we have presented numerical evidence for a strongly anomalous non-thermal fixed point in a two-dimensional Bose gas.
Starting from different kinds of ensembles of elementary and non-elementary vortices, \ie{}, from a far-from-equilibrium initial state, the evolving isolated system shows a slow decay of the vortices through mutual annihilation.
The decay can be described as coarsening dynamics, showing an anomalously slow self-similar evolution of the single-particle momentum spectrum, Eqs.~\eq{NTFPScaling} and \eq{nkitofskappa}, characterized 
by scaling exponents, $\alpha=0.402\pm0.05$, $\beta=0.193\pm0.05$, and $\zeta = 5.7\pm0.3$, see  Eqs.~\eq{82} and \eq{83}, Figs.~\fig{vorlat_spectrum_unresc} and \fig{vorlat_spectrum}, implying a dynamical scaling exponent $z_{\text d}\simeq5$.
This anomalous non-thermal fixed point is reached prior to a different, (near-)Gaussian fixed point exhibiting exponents $\alpha=1.10\pm0.08$,
$\beta=0.56\pm0.08$, and $\zeta=4.0\pm0.1$, see Eqs.~\eq{alphabetag} and \eq{zetagaussian}, as well as Figs.~\fig{ranvort_spectrum_unresc} and \fig{ranvort_spectrum}. 

We demonstrate that the self-similar coarsening evolution of the isolated system near the Gaussian fixed point is consistent with the dissipative coarsening when the system is coupled to a thermal bath.
This is attributed to the fact that also the isolated system builds up thermal-like fluctuations at short wave lengths.
It is a matter of the strength of this coupling to the bath when the Gaussian fixed-point scaling starts to supersede the anomalous one. 
As a consequence, the observed coarsening represents a type of dynamical critical behavior which appears to be closely related to the equilibrium phase transition at non-vanishing temperatures.
In contrast to the entirely thermal state, however, it shows a steep Porod law, \Eq{zetagaussian}, associated with randomly scattered elementary vortices and anti-vortices on a phase-coherent background \cite{Nowak:2011sk,Nowak:2012gd} which are bound to decay.

On the other hand, provided the system is prepared in a suitable initial state, allowing for clustering of like-sign vortices, it exhibits the strongly anomalous scaling evolution  at early times.
The associated anomalous scaling exponent $\eta\simeq -3$, \cf{}~\Eq{etaanomalous}, is found to take the same role as the exponent $\mu$ characterizing the conservation law governing the coarsening dynamics of the order parameter in phase-ordering kinetics.

Our results indicate that, far from equilibrium, the model possesses a non-thermal fixed point seemingly unrelated to the thermal fixed point.
The respective anomalous scaling is seen to last as long as the coupling to thermal noise is sufficiently weak such that three-body collisions of defects represent the dominant loss process.
The clustering is reflected in a steeper Porod law at low wave numbers, \Eq{83}, which, as for the Gaussian fixed point, disappears in the thermal state. 
In both cases, the critical behaviour is possible only through strong driving out of equilibrium as accomplished by the initial quench which leads to the coherent hydrodynamic propagation of vortices on the background of a strongly phase-coherent gas.

% ==============================================================================
% ==============================================================================
\section*{Acknowledgments}
The authors thank J. Berges, T. Billam, F. Brock, I. Chantesana, S. Diehl, S. Erne, C. Ewerz, A. Fr\"olian, M. Kastner, P. Kunkel, D. Linnemann, J. Marino, S. Mathey, T. M\"uller, M. K. Oberthaler, J. M. Pawlowski,  A. Pi{\~n}eiro Orioli,  M. Pr\"ufer, H. Salman, A. Samberg, M. Schiffer, C. Schmied, and H. Strobel for discussions and collaboration on the topics described here. 
This work was supported by the Horizon-2020 framework programme of the EU (FET-Proactive, AQuS, No. 640800),  by Deutsche Forschungsgemeinschaft (SFB 1225 ISOQUANT and Grant No.~GA677/8), by the Helmholtz Association (HA216/EMMI), and by Heidelberg University (Center for Quantum Dynamics).\\

%======================================================================================
%======================================================================================
%\clearpage
\appendix
\begin{center}
{\bf APPENDIX}  
\end{center}
\vspace{-3mm}
\numberwithin{equation}{section}
\numberwithin{figure}{section}
\setcounter{section}{0}
\setcounter{equation}{0}
\setcounter{figure}{0}

%======================================================================================
%======================================================================================
\section{Dynamical simulations}
\label{app:DynamicalSimulations}
In this paper we statistically simulate the far-from-equilibrium
dynamics in the classical-wave limit of the underlying quantum field
theory. The classical equation of motion for the complex scalar field
$\phi(\vector{x},t)$ reads
\begin{equation} 
\label{eq:GPE} \i \del_t \phi(\vec{x},t)= \left[
-\frac{\nabla^2}{2m}+g|\phi(\vec{x},t)|^2 \right] \phi(\vec{x},t) .
\end{equation}
Here, $m$ is the boson mass, $g$ quantifies the interaction strength
in $d=2$ dimensions, and, in our units, $\hbar=1$.  Our computations
are performed in a computational box of size $L^2$ on a grid with side
length $L=N_s a_s$, lattice spacing $a_s$, and periodic boundary
conditions.
We implement a split-step integration scheme on  
NVIDIA \emph{GTX Titan Black} GPUs, in a cluster with 4 host nodes and 4 graphic cards per node.
For this, the core functions of the propagation scheme, in particular the fast Fourier transform, 
are implemented using NVIDIA's CUDA extension for C \cite{Nickolls2008a}.

For the numerical data presented in this work, we solve \Eq{GPE} on a
grid with $N_s = 1024$ interpolation points and lattice spacing
$a_s= 1$ in each direction. Therefore, the only length scale which
remains to be fixed is the healing length
$\xi_h = (2mgN/L^2)^{-1/2} = (2mgN/N_s^2)^{-1/2}\,a_s$. We choose a mass $m=1/2$, 
a total number of particles $N = 3.2 \cdot 10^9$, an interaction strength
$g = 3 \cdot 10^{-5}$. This results in a healing length of
$\xi_h = 3.30\, a_s$, which determines the resolution of the core of an 
elementary vortex on the computational grid.
Taking, as an example, the 2D experiment in Ref.~\cite{Desbuquois2012a.NatPhys.8.645}, the elementary time step corresponds to $2m\xi_{h}^{2}/\hbar=\xi_{h}/c_{s}\simeq0.2\,$ms, such that a total time of $10^{5}\,\xi_{h}^{2}$ corresponds to about $20$ seconds.

%======================================================================================
\section{Time evolution of spatial defect patterns}
\label{app:SpatialPatterns}
\begin{figure}[!t]
  \includegraphics[]{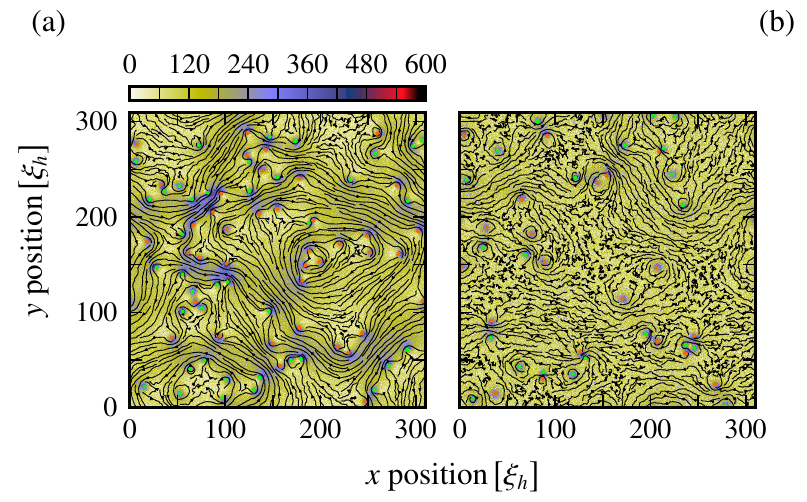}
  \caption{%
    The hydrodynamic velocity field $\vec{v}$ corresponding to the
    snapshots of the time evolving density shown in \Fig{vorlat}
    (panel a).  Black flow lines indicate the orientation of the
    velocity field while the colour projection depicts the modulus
    $\lvert \vec{v} \rvert$. The positions of vortices (anti-vortices)
    are marked by orange (green) dots.  The strong clustering after
    the initial break-up as well as the large-scale vortex clustering
    leading to strong coherent flows during the universal, late-time
    stage are clearly seen. Panel b shows a characteristic
    hydrodynamic velocity field near the Gaussian non-thermal fixed
    point. Here, the system largely consists of relatively closely
    bound pairs of vortices and anti-vortices. Both flow fields are
    depicted at a time $t=4\cdot 10^4\,\xi_h^2$.}
  \label{fig:flowlinesComparison}
\end{figure}

In this appendix we complement \Fig{flowlines} in \Sect{SpatialPatternEvolution} with a diagram comparing the flow patterns during the dynamics near the anomalous fixed point with that near the Gaussian fixed point.
\Fig{flowlinesComparison} shows, in panel (a), the hydrodynamic velocity field $\vec{v}$ at time $t=4\cdot 10^4\,\xi_h^{2}$ during the evolution near the anomalous fixed point.
Black flow lines indicate the orientation of the velocity field while the color projection depicts the modulus $\lvert \vec{v} \rvert$. 
It exhibits clustering of vortices (orange dots) and anti-vortices (green dots) which induce strong coherent flows indicated by colors on the blue side of the scale.
The clustering leads to a screening of the vortices enclosed in each cluster against mutual annihilation with vortices of the opposite circulation.
Decay of the vortex number can effectively be ascribed to three-vortex-collision loss, as described in \Sect{ThreeBodyKinetics}.

This configuration is compared, in \Fig{flowlinesComparison}b, to a typical configuration at at time $t=4\cdot 10^4\,\xi_h^{2}$ during the evolution near the Gaussian fixed point.
Here, the system largely consists of relatively closely bound pairs of vortices and anti-vortices.
The scattering of these pairs leads to a decay of the total number of defects which scales like $N_{\text d}\sim t^{-1}$, see \Sect{ThreeBodyKinetics}.

%======================================================================================
\section{Coarsening in the presence of a thermal bath}
\label{app:ThermalCoarsening}
In this appendix, we discuss the coarsening dynamics of a 
driven-dissipative system, \ie{}, a Bose gas in a thermal environment, 
for comparison with the universal dynamics
near non-thermal fixed points in the isolated system.
Instead of sampling over stochastic \emph{initial} field distributions 
a time-dependent stochastic driving force and dissipation via imaginary 
coupling constants are introduced~\cite{Cockburn2011a,2012PhRvA..86a3627G,Cockburn2012a}.
We briefly recall the driven-dissipative stochastic Gross-Pitaevskii equation.
which we then use to determine numerically the coarsening dynamics, \ie{}, 
the coarsening exponents $\alpha$ and $\beta$.

%==================================================
%
%==================================================

%======================================================================================
\subsection{Driven-dissipative Gross-Pitaevskii equation}
\label{app:DDGPE}
The dilute interacting Bose gas coupled to a particle
bath and a driving force field is described, in semi-classical approximation, by the stochastic
driven-dissipative GPE~\cite{Cockburn2012a}
\begin{equation}
\label{eq:sGPE}
i \partial_t\phi(\vec{x},t) = \frac{\delta H}{\delta\phi^{\ast}(\vec{x},t)} 
+ \frac{\delta H_{\text{d}}}{\delta\phi^{\ast}(\vec{x},t)} 
+\zeta(\vec{x},t)\, ,
\end{equation}
where $\zeta$ is a stochastic external force vanishing in the mean, $\langle \zeta(\vec{x},t) \rangle \equiv 0$, and ${H}$ denotes the standard Gross-Pitaevskii Hamiltonian,   
\begin{equation}
  \label{eq:GPEreal}
  \frac{\delta H}{\delta\phi^{\ast}(\vec{x},t)} 
  = \left[-\frac{1}{2m}\nabla^2 + g\left\lvert\phi(\vec{x},t)\right\rvert^2 -\mu\right]\phi(\vec{x},t)\,,
\end{equation}
with particle mass $m$ and chemical potential  $\mu$.  $g$ is the real part
of the coupling constant, defined, in $d=2$ dimensions by $g = - (4 \pi / m)[\ln(\mu m a^2/4)]^{-1}$
in terms of the $s$-wave scattering length $a$.  
$H$ describes the energy and particle-number conserving
part of the dynamics, while  $H_{\text{d}}$ accounts for the dissipative part of the
dynamics,
\begin{equation}
  \label{eq:GPEdiss}
  \frac{\delta H_{\text{d}}}{\delta\phi^{\ast}(\vec{x},t)} 
  = i \left[\nu\nabla^2-g_\text{d}\left\lvert\phi(\vec{x},t)\right\rvert^2 +\mu_\text{d}\right]\phi(\vec{x},t)\,,
\end{equation}
defined in terms of the diffusion constant $\nu$ and the imaginary
parts of the coupling, $g_{d}$, and chemical potential, $\mu_{d}$.  
We restrict ourselves to the case
$g_\text{d}>0$ and $\mu_\text{d} >
0$ for which the terms proportional to $g_\text{d}$ and $\mu_\text{d}$
account for loss and gain of particles, respectively. 

The stochastic force $\zeta$ is chosen to have Gaussian statistics, 
uncorrelated in time and space,
\begin{equation}
\langle \zeta^{\ast}(\vec{k},t)\zeta(\vec{k},t') \rangle = {\gamma} \,\delta(t-t')\, ,
\label{eq:correlator}
\end{equation}
with a 
driving strength $\gamma$. 
In our simulations, $\gamma$ is a constant in time and space.  

While, in general, the above setup generates a strong non-equilibrium coupling to a
non-thermal bath, a special choice of the parameter set
renders the setup thermal~\cite{Cockburn2012a}. 
The complex coupling constants $1/2m + \i \nu$, $g+\i g_\text{d}$, and $\mu + \i
\mu_{\text{d}}$ need to have the same argument in order for the system to
obey detailed balance~\cite{Sieberer2013a}, \ie{}, $\delta=g_\text{d}/g=\mu_\text{d}/\mu=2m\nu$. 
As a result of the classical-statistical fluctuation-dissipation relation, the temperature is related to this fraction $\delta$ by~\cite{Cockburn2012a} 
\begin{equation}
  \label{eq:Titoalphagamma}
 T = 2k_\text{B}\delta/\gamma\,. 
\end{equation}
%

%======================================================================================
\subsection{Coarsening dynamics}
\label{app:thermal}
We use the stochastic Gross-Pitaevskii approach summarized above to
simulate coarsening dynamics of the 
dilute Bose gas coupled to a thermal bath for comparison with
our studies of the isolated system.
We choose $H_\text{d} = i\delta\, H$ in \Eq{sGPE}
otherwise keeping the parameters chosen in the main text.
The imaginary parts of the coupling parameters are chosen such that 
$\delta=0.00215$, and $\gamma$ is adjusted to yield 
temperatures $T= 100\,\xi_h^{-2}$.
This places the system deep in the quasi-ordered regime below the
Berezinskii--Kosterlitz--Thouless transition temperature which, in mean-field
approximation, is $T_{\text{BKT}}=\pi \sqrt{n_0}/2m=574\xi_h^{-2}$~\cite{Wen2004a.QFT}.
As a consequence, no free vortices are possible once the solutions have 
reached the stationary state.

%==========================================
\begin{figure}[!t]
\centering
\includegraphics{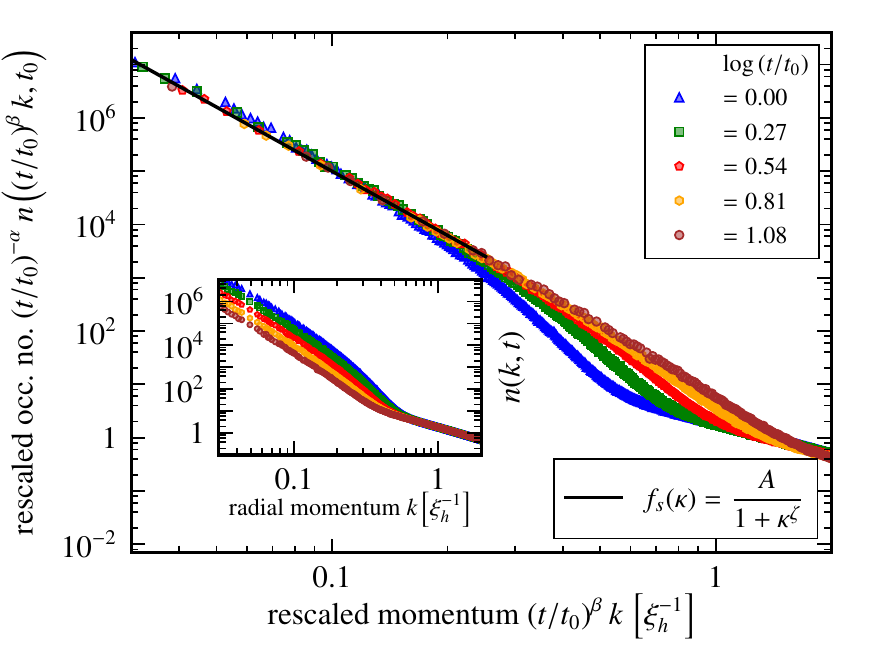}
\includegraphics{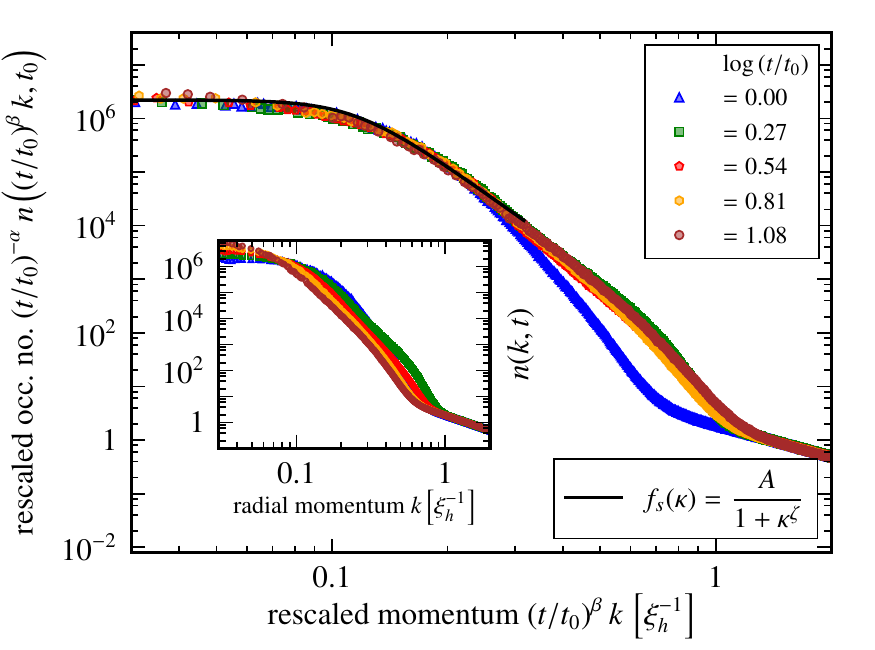}
\caption{Scaling analysis of the time evolving single-particle spectra
  for the vortex lattice initial condition with $8 \time 8$ vortices
  of winding number $w = \pm6$, when coupled to a thermal bath with
  $\delta = 0.00215$ at temperature $T = 100\xi_h^{-2}$. \emph{Upper
    panel}: Scaling collapse at late times, in the interval $1\cdot
  10^4 \xi_h^2 < t < 1\cdot 10^5 \xi_h^2$.  The inset shows the
  non-rescaled spectra at five different, logarithmically equidistant
  times as given in the legend.  The times are defined relative to the
  reference time $t_0 = 1\cdot10^4\,\xi_h^2$.  The main graph depicts
  the scaling collapse of the spectrum according to $n(t,k) =
  (t/t_0)^{-\alpha}n([t/t_0]^{\,\beta} k,t_{0})$.  The corresponding
  scaling exponents, determined from a fit of $100$ logarithmically
  equidistant times, are $\alpha = 1.2 \pm 0.05$ and $\beta= 0.53
  \pm0.05$.  The universal scaling function in this time interval does
  not display the cut-off of the scaling form $k_\lambda$ any more, as
  it is beyond the infrared cut-off of the computational
  grid. However, we determine a scaling exponent $\zeta = 4.0\pm 0.1$
  from fitting a power law to the infrared part of the rescaled
  spectra (cf.~\App{fitting}).  \emph{Lower panel}: Same analysis as above for the time
  interval $2\cdot 10^2 \xi_h^2 < t < 2\cdot 10^3 \xi_h^2$. The
  corresponding scaling exponents, determined from a fit of $100$
  logarithmically equidistant times, are $\alpha = 0.41 \pm 0.05$ and
  $\beta= 0.22 \pm0.05$. The universal scaling form in this time
  interval is consistent with that near the anomalous fixed point in
  the isolated system, with a scaling exponent $\zeta = 5.7\pm0.1$.}
  \label{fig:thermal_spectrum}
\end{figure}
%==========================================
We use this setup to compute the dynamics of the occupation-number
spectrum in analogy to the evolution of the isolated system,
\cf{}~\Sect{NTFP}.  \Fig{thermal_spectrum} shows the approach to the
thermal state from a vortex-lattice initial state with initially $8
\times 8$ vortices of winding number $w = \pm 6$.  For this parameter
setting, the time evolution of the mean vortex distance, $\ell_{\text d}(t)$,
after coupling the system to the thermal bath, is shown in
\Fig{ThermalCoarsening} (\cf{}~red data points).

The coarsening evolution of the defect scale shows two distinct
scaling regimes, for which we analyse the time evolution of the
occupation spectrum separately. We find that the occupation spectrum
can be rescaled to a universal curve at early times, in the time
interval $2\cdot 10^2 \xi_h^2 < t < 2\cdot 10^3 \xi_h^2$, and at late
times, $10^4 \xi_h^2 < t < 10^5 \xi_h^2$, before all
defects have decayed and the spectrum eventually indicates
thermalisation.

The scaling collapse of the distribution at $100$ different times in the interval
$10^4\,\xi_h^2<t<10^5\,\xi_h^2$
yields the scaling exponents (see~\App{fitting})
\begin{align}
  \label{eq:88a}
  \nonumber  
  \alpha^{\text{th}}_{\text g} 
  &= 1.2 \pm 0.05\,,
  \\
  \beta^{\text{th}}_{\text g} 
  &= 0.53 \pm0.05\,.
\end{align}
In the upper panel of \Fig{thermal_spectrum}, we show the rescaled spectra at five exemplary times within the above interval, together with the non-rescaled spectra in the inset.
These exponents are consistent with those near
the Gaussian non-thermal fixed point approached in the isolated system, \cf{}~\Eq{alphabetag}. 
The infrared momentum power-law exponent $\zeta=4.0\pm 0.1$ 
is consistent with the Porod law \eq{zetagaussian} describing the 
random vortex distribution near the Gaussian fixed point. 

Remarkably, we also find the anomalously slow self-similar evolution
of the gas coupled to a thermal bath at early times, $2\cdot 10^2
\xi_h^2 < t < 2\cdot 10^3 \xi_h^2$, inducing strong vortex clustering
during the initial stage of the evolution.  In this case we observe
strongly anomalous coarsening, with the exponents
\begin{align}
  \label{eq:88b}
  \nonumber  
  \alpha^{\text{th}}_{\text a} 
  &= 0.41 \pm 0.05\,,
  \\
  \beta^{\text{th}}_{\text a} 
  &= 0.22 \pm 0.05\,.
\end{align}
To obtain these values from a fit, we used spectra at $100$
logarithmically equidistant times in the interval $2\cdot 10^2 \xi_h^2
< t < 2\cdot 10^3 \xi_h^2$. Examples for the rescaled distributions
are depicted in the lower panel of \Fig{thermal_spectrum} at five
exemplary times. The collapsed spectra follow a universal scaling
function $f_s(\kappa)$ with an infrared scaling exponent $\zeta = 5.7
\pm 0.1$. Thus, at early times, the set of scaling exponents is
consistent with the scaling at the anomalous fixed point in the isolated system.

Note that the thermally coupled system displays the two types of fixed-point evolution for all values of $\delta$ analysed in
\Fig{ThermalCoarsening}.  As discussed in the main text, the coupling
strength to the thermal bath changes the characteristic time
$t_\text{c}$ at which the scaling properties change from anomalous to Gaussian \footnoterecall{fn:Video}. 

%======================================================================================
\subsection{Compressible vs. imcompressible excitations}
\label{app:CompVsIncompSpectra}
%
%==========================================
\begin{figure}[!t]
\centering
    \includegraphics{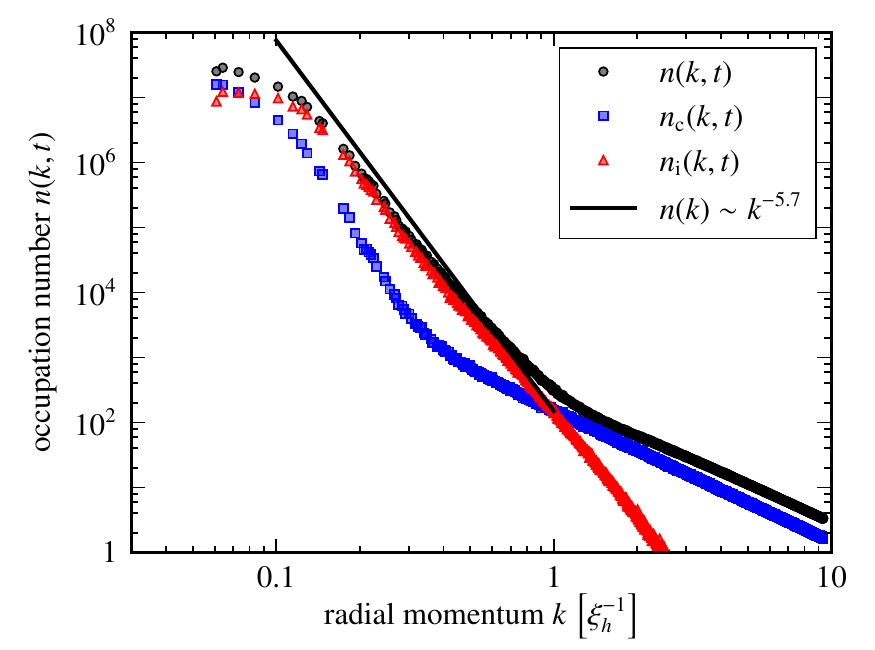}
    \includegraphics{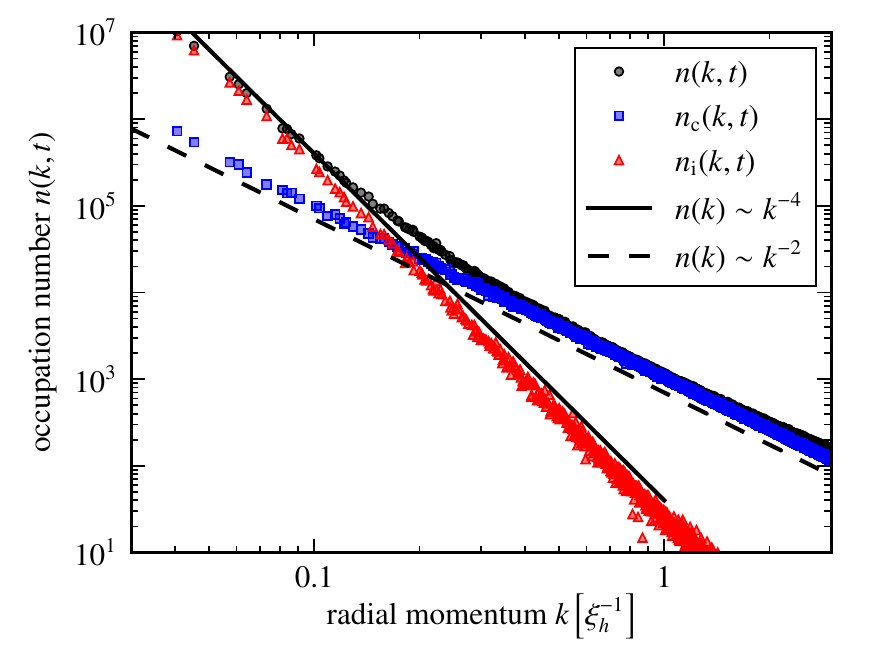}
    \caption{Momentum distributions of the occupation numbers $n_\mathrm{i}(k)$ and $n_\mathrm{c}(k)$, together with the total single-particle spectrum $n(k)$, each at a late time when the system is close to the fixed point.
Red circles denote $n_{\mathrm{i}}$, filled blue squares $n_{\mathrm{c}}$. In both figures, the decomposition is depicted at a time $t=4\cdot 10^5\,\xi_h^2$.
While the upper panel shows the occupation numbers close to the anomalous fixed point, the lower panel relates to the Gaussian fixed point.
The anomalous fixed point is characterized by a much larger compressible contribution in the regime of a steep Porod law.
    }
  \label{fig:CompVsIncompSpectra}
\end{figure}
%==========================================
In this appendix we recapitulate the decomposition of the superfluid flow pattern in the system near the non-thermal fixed point, into transverse (incompressible) and longitudinal (compressible) contributions. 
In this way we can show that the IR scaling is dominated by the incompressible part while in the UV the particles mainly belong to the compressible as well as a quantum pressure components. 

To define the decomposition we use the polar representation $\phi(\mathbf{x},t)=\sqrt{\rho(\mathbf{x},t)}\exp\{i\theta(\mathbf{x},t)\}$ of the field in terms of the density $\rho(\mathbf{x},t)$ and a phase angle $\theta(\mathbf{x},t)$. 
This allows to express the particle current $\mathbf{j}=i(\phi^*\nabla \phi - \phi \nabla \phi^*)/2=\rho\mathbf{v}$ in terms of the velocity field $\mathbf{v}=\nabla\theta$.

We decompose the kinetic-energy spectrum $E_{\mathrm{kin}}= \int \mathrm{d}^dx \, \langle |\nabla \phi(\mathbf{x},t)|^2\rangle/(2m)$ following \cite{Nore1997a}  as $E_{\mathrm{kin}} = E_{\mathrm{v}} + E_\mathrm{q}$ into a `classical' part 
\begin{equation}
\label{eq:Ev}
E_\mathrm{v}= \frac{1}{2m}\int \mathrm{d}^dx \, \langle |\!\sqrt{\rho}\mathbf{v}|^2 \rangle 
\end{equation}
and a `quantum-pressure' component 
\begin{equation}
E_\mathrm{q}=\frac{1}{2m}\int \mathrm{d}^dx \, \langle |\nabla\!\sqrt{\rho}|^2 \rangle \,.
\end{equation}
The radial energy spectra for these fractions involve the Fourier transform of the generalized velocities $\vector{w}_{\mathrm{v}}=\sqrt{\rho}\vector{v}$ and $\vector{w}_{\mathrm{q}}=\nabla\sqrt{\rho}$,
\begin{eqnarray}
 E_{\delta}(k)= \frac{1}{2m} \int k^{d-1} \mathrm{d}\Omega_d \, \langle |\mathbf{w}_{\delta}(\mathbf{k})|^2 \rangle,\quad \delta=\mathrm{v},\mathrm{q}.
\end{eqnarray}
Following Ref.~\cite{Nore1997a}, the velocity $\mathbf{w}_{\mathrm{v}}$, which due to the multiplication of $\mathbf{v}$ with the density $\rho$ becomes regularized and can be Fourier transformed, is furthermore decomposed into `incompressible' (divergence free) and `compressible' (solenoidal) parts, $\mathbf{w}_{\mathrm{v}}=\mathbf{w}_{\mathrm{i}}+\mathbf{w}_{\mathrm{c}}$, with $\nabla \cdot \mathbf{w}_\mathrm{i}=0$, $\nabla \times \mathbf{w}_\mathrm{c}=0$, to distinguish vortical superfluid and rotationless motion of the fluid. For comparison of the kinetic-energy spectrum with the single-particle spectra $n(k)$, we determine occupation numbers corresponding to the different energy fractions as 
\begin{equation} 
\label{eq:decomposition}
n_\mathrm{\delta}(k) =  k^{-d-1}E_{\mathrm{\delta}}(k)\,,\,\, \delta \in \{\mathrm{i}, \mathrm{c}, \mathrm{q}\}. 
\end{equation}
The resulting spectra $n_\mathrm{i}(k)$, $n_\mathrm{c}(k)$, and $n_\mathrm{q}(k)$ add up to $n_\mathrm{s}(k) = n_\mathrm{i}(k) + n_\mathrm{c}(k) + n_\mathrm{q}(k)$, which agrees with the single-particle spectrum up to small corrections \cite{Nowak:2011sk}.

In \Fig{CompVsIncompSpectra}, we depict the momentum distributions of the occupation numbers $n_\mathrm{i}(k)$ and $n_\mathrm{c}(k)$, together with the total single-particle spectrum $n(k)$, each at a late time when the system is close to the fixed point.
Red circles denote $n_{\mathrm{i}}$, filled blue squares $n_{\mathrm{c}}$.
While the upper panel shows the occupation numbers close to the anomalous fixed point, the lower panel relates to the Gaussian fixed point. 

%======================================================================================
\section{Fitting procedure}
\label{app:fitting}
In this appendix, we discuss the numerical fitting procedure employed
in determining the scaling exponents $\alpha$ and $\beta$ which characterize
the self-similar time evolution of the occupation
spectrum. Furthermore, we comment on the fits of the universal scaling
functions, leading to the spatial scaling exponents $\zeta$. The
procedure explained here is used for the data sets presented in
Figs.~\ref{fig:vorlat_spectrum}, \ref{fig:ranvort_spectrum}, and \ref{fig:thermal_spectrum}
and yields the sets of exponents given in Eqs.~\eq{82}, \eq{alphabetag}, \eq{88a}, and \eq{88b}.

The basic strategy for obtaining the temporal scaling exponents is as
follows: First, we select a suitable time window within the universal
regime, by inspecting the time evolution of a characteristic scale
such as $\ell_{\text{d}}$. Within this window, we compute the
occupation spectrum $n(k,t_j)$ at several times $t_j$,
where the times $t_j$ are chosen with a logarithmically equidistant
spacing. Then, the distance between a chosen reference spectrum
$n(k,t_\text{ref})$ and all rescaled spectra
$(t_j/t_\text{ref})^{-\alpha}n([t_j/t_\text{ref}]^{-\beta}k,t_{j})$ within the window is
numerically minimized with respect to $\alpha$ and $\beta$.

As the occupation spectra are obtained by numerically
solving the Gross--Pitaevskii equation on a discrete grid, the spectra
$n(k,t_j)$ at each time step are available for fixed, discrete
radial momenta $k$ only. Therefore, to be able to evaluate the
rescaled spectrum $(t_j/t_l)^{-\alpha}n([t_j/t_l]^{-\beta}k,t_{j})$,
the data is interpolated by means of cubic B-splines, 
as provided by the \textit{Python-SciPy} library~\cite{scipy}.

In the above procedure, we minimize the quantity 
\begin{align}
  \label{eq:chisquare}
  \nonumber &\chi(\alpha, \beta) 
  =\\ 
  &\quad \sum_{k<k_\Lambda,j} \Big\{\log{n(k,t_\text{ref})} 
    - \log{\big[(t_j/t_\text{ref})^{-\alpha}n([t_j/t_\text{ref}]^{-\beta}k,t_{j})\big]}\Big\}^{2} 
\end{align}
summing over all discrete momenta $k$ up to a UV
cut-off $k_\Lambda$ and over all rescaled spectra within the scaling
window. The cut-off $k_\Lambda$ is set by hand, to exclude the UV tail
of any of the spectra from the rescaling procedure. Anticipating
positive exponents $\alpha$ and $\beta$, this can be achieved by
determining $k_\Lambda$ based on the spectrum $n(k,t_\text{max})$ at the
latest time $t_\text{max}$ within the scaling window.

Finally, to minimize the distance \eq{chisquare} for a set of
occupation number spectra, we employ the Levenberg--Marquardt algorithm, as
provided by the \textit{Python-SciPy} library~\cite{scipy}. This yields
the values for $\alpha_\text{min}$ and $\beta_\text{min}$ for a chosen
reference spectrum and a set of times $t_j$. The errors for these values
are estimated using the numerically constructed Jacobian of $\chi$
(as provided by the \textit{SciPy} implementation of the algorithm) at
the minimum and the residual variance $\chi(\alpha_\text{min},\beta_\text{min})$.

To further assess the robustness of the fit, we additionally vary the
reference time $t_\text{ref}$ and the sample times $t_j$ which are used in the
fitting procedure for a specific scaling window. The values for
$\alpha$ and $\beta$ stated in the main text, together with their
errors, result from averaging $\alpha_\text{min}$ and
$\beta_\text{min}$ from the Levenberg--Marquardt fits over these
additional variations.

Once the optimal values for $\alpha$ and $\beta$ are determined as
described above, the occupation spectra interpolated within the
scaling window can be collapsed by rescaling and be averaged. This yields
a numerical estimate for the scaling function $f_s(\kappa)$. To this
estimate, we fit the ansatz \eq{4} for the scaling function, again
using the Levenberg--Marquardt algorithm on the distance of
logarithms. The normalization constant $A$ is fixed by requiring
$\lim_{\kappa \to 0}f_s(\kappa) = 1$. The exponent $\zeta$ and the
momentum scale $k_\lambda$ are treated as open fitting parameters,
whereas the UV cut off $k_\Lambda$ is fixed by hand.

%\bibliography{Master}
%\bibliography{Master_MKarl,Master}
%\bibliography{Master_MKarl,Bibliography/Master}

%\clearpage
%

\end{document}